\documentclass[twocolumn,showpacs,preprintnumbers,amsmath,amssymb,nofootinbib]{revtex4-1}
\usepackage{graphicx}
\usepackage{dcolumn}
\usepackage{bm}

\usepackage{pstricks}
\usepackage{psboxit}
\usepackage{color}

\begin{document}
\newcommand{\D}{{\rm d}}

\title{Equilibration in the time-dependent Hartree-Fock approach
probed with the Wigner distribution function}
\author{ N. Loebl$^1$, J.A. Maruhn$^1$, and P.-G. Reinhard$^2$}
 
\affiliation{
$^1$Institut fuer Theoretische Physik, 
Universitaet Frankfurt, D-60438 Frankfurt, Germany 
} 
\affiliation{
$^2$Institut fuer Theoretische Physik II,
Universitaet Erlangen-Nuernberg, D-91058 Erlangen, Germany
}
\date{\today}

\begin{abstract}
Calculating the Wigner distribution function in the reaction plane, we
are able to probe the phase-space behavior in time-dependent
Hartree-Fock during a heavy-ion collision. We compare the Wigner
distribution function with the smoothed Husimi distribution function.
Observables are defined to give a quantitative measure for local and global
equilibration. We present different reaction scenarios by analyzing
central and non-central $^{16}O+$$^{16}O$ and $^{96}Zr+$$^{132}Zn$
collisions. It is shown that the initial phase-space volumes of the
fragments barely merge. The mean values of the observables
are conserved in fusion reactions and indicate a "memory effect" in
time-dependent Hartree-Fock. We observe strong dissipation but no
evidence for complete equilibration.
\end{abstract}

\pacs{21.60.-n,21.60.Jz}

\maketitle

\section{Introduction}

The time-dependent Hartree-Fock (TDHF) method was originally proposed as
early as 1930 by Dirac~\cite{Dirac-tdhf}.  For a long time, it was merely a
formal tool to derive many-body approaches like, e.g., in \cite{Bro71aB} to
derive linear response theory. The enormous progress of computational
facilities has made TDHF a practical scheme for describing the dynamics
of many-body systems.  By now it has found widespread applications in
various areas of physics.  Under the label of time-dependent density-functional
theory it is used in electronic systems like atoms, molecules, clusters, and
solids, see e.g. \cite{Dreizler,Rei03a,tddft-notes}.  The earliest practical
applications probably appeared in nuclear physics \cite{Bonche76}, where TDHF
is a powerful microscopic approach to simulate various dynamical scenarios in
the regime of large-amplitude collective motion, like fusion excitation
functions, fission, deep-inelastic scattering, and collective excitations; for
early reviews see, e.g., \cite{Svenne,Negele,Davies}.
These pioneering applications were still hampered by the 
computational limitations of their time. 
With the ongoing growth of computational power, fully three-dimensional TDHF
calculations without any symmetry restriction became feasible and so
renewed the interest in nuclear TDHF, for a few recent examples of
state-of-the-art TDHF calculations in many different processes see
\cite{Kim,Simenel,Nakatsukasa,Umar05a,Maruhn1,Guo08a}.

The TDHF approach allows the self-consistent
quantum-mechanical description of nuclear dynamics on a mean-field level.
Self-consistency means an unprejudiced description once a reliable energy
functional is given. This explains the versatility of TDHF. It
remains, however, an approximation 
since it is a mean-field theory. TDHF misses dynamical
correlation effects stemming from nucleon-nucleon collisions, which contribute
to (two-body) dissipation and thermalization. Their inclusion in a fully
quantum mechanical treatment has so far only been achieved in homogeneous
systems like, e.g., \cite{Toe88a,Gre94a}.  Including dynamical correlations
for finite nuclei is presently still restricted to a semiclassical
description \cite{Bertsch,Bonasera, Abe}. On the other hand, it was found that
nuclear TDHF calculations already include a great deal of (one-body) dissipation
if all terms of the functional, particularly the spin-orbit terms, are
properly accounted for \cite{Rei88d} and if all symmetry restrictions are
removed \cite{Mar06c}. This dissipation within TDHF does not result from
two-particle collisions but from collision of one particle with the boundaries of
the moving mean-field potential (``single-particle dissipation''
\cite{Swiatecki}) which randomizes the single-particle states. In a heavy-ion
collision, two pictures of single-particle dissipation can be
distinguished. The ``window'' picture describes dissipation of relative momentum
via nucleon exchange through a neck while the ``wall'' picture deals with the
dissipation of kinetic energy by reflection of the nucleons at a moving wall
\cite{Blocki,Randrup1,Randrup2}. The latter results in a net increase of the
nucleons' thermal energy provided there is no correlation between the nucleonic
and wall motions. However, these are idealized concepts which are not always
immediately applicable to realistic heavy-ion collisions
\cite{Maruhn2,Sierk,Koonin}. Until now it is not understood at a detailed
level how rapidly and how strongly equilibration works within the TDHF
approach.

A rough global measure of dissipation is given by comparing initial and final
kinetic energies of the fragments in a heavy-ion collision \cite{Mar06c}. More
detailed analysis should look at something like a local momentum
distribution. This naturally leads to the concept of a Wigner function which
provides a phase-space picture of a quantum state. Originally introduced in
\cite{Wigner}, it is often used for establishing the connection between quantum
and classical physics \cite{Bra97aB}.  The result of such semiclassical
limits is a mean-field dynamics in classical phase-space called the Vlasov
equation \cite{Vlasov} which is widely used in simulating nuclear dynamics
\cite{Bertsch,Bonasera,Abe}.  In this paper, we want to stay at the fully
quantum-mechanical level and employ the Wigner function as a useful
observable helping to analyze TDHF dynamics.  An early analysis of that kind
is found in \cite{Maruhn3}. The Wigner function has the weakness that it is
not positive semidefinite, thus preventing a strict probabilistic
interpretation. This defect is cured by some phase-space smoothing leading to
the Husimi function \cite{Takahashi,Toscano}, which also turns out to be the
better starting point for the semiclassical expansion \cite{Eplattenier}. We
will also briefly address the Husimi function in connection with TDHF results.
As the Wigner function is six-dimensional and thus rather difficult to handle, 
we deduce from it more
compact measures of dissipation and equilibration by considering local
quantities integrated with some weights over momentum space, e.g.,
the eccentricity of the momentum space distribution. These observables are
complemented by others computed without recurring to the
Wigner picture, e.g., the intrinsic excitation energy which is computed
from the local kinetic energy density. We will explore these different
analyzing tools for two realistic applications, collision of $^{16}$O+$^{16}$O
and $^{96}$Zr+$^{132}$Sn.

The paper is organized as follows: Section \ref{sec:numeric} describes briefly
the numerical handling of TDHF used in this work.  In Section \ref{sec:Wigner}
we present the transformation from the TDHF wave function to the Wigner and
Husimi representations. Results for the ground states in static calculations
are presented to compare both pictures.  Observables are defined in Section
\ref{sec:observ} to allow a quantitative discussion of equilibration. In
Section \ref{sec:results} we show results for dynamical calculations with
different nuclei, energies, and impact parameters.

For sake of generality the formal considerations of Section
\ref{sec:Wigner} are presented in $n$-dimensional coordinate
and $2n$-dimensional phase space. The results in this paper are
obtained in the reduced two-dimensional reaction plane
(assumed to be the $x$-$z$-plane). For clarity we will
label the number of coordinate dimensions $n$ of the applied
distribution function $f$ with $f^{(n)}$. 

\section{Formal and numerical framework}
\label{sec:numeric}

The basis of the TDHF description is a set of occupied single-particle
wave functions $\psi_l(\mathbf{r},t)$ where $l$ labels the
states. These wave functions are two-component spinors. 
The Skyrme mean-field Hamiltonian is computed for given densities and
currents in the standard manner \cite{Ben03aR}.
 For all calculations reported here we
have used the Skyrme parametrization SkI3 \cite{Reinhard2}.

The TDHF equations are solved on a three-dimensional Cartesian
coordinate-space grid. Using the fast Fourier transformation (FFT) derivatives
can be evaluated very efficiently in Fourier-space. The mesh spacing is 
${\rm d}x={\rm d}y={\rm d}z=1$\:fm. 

The stationary ground states of the initial systems are computed via the
damped-gradient iteration algorithm \cite{Blum,Reinhard}. The initial state is
obtained by placing the ground states of the two fragments in a safe distance
and giving them a boost towards each other.  These states are then propagated
in time by use of a Taylor-series expansion of the time-evolution operator
\cite{Flocard} where the expansion is taken up to sixth order.  The actual
time step is $t=0.2$\:fm/c.

\section{Wigner and Husimi distributions}
\label{sec:Wigner}

The Wigner function is a transformation of the density matrix to a phase-space
function. There are various levels of density matrices in a many-body systems
and accordingly various Wigner functions.  TDHF can be considered as
describing the dynamics of the one-body density matrix
$\rho(\mathbf{r},\mathbf{r}')$, neglecting all correlations
between the interacting nucleons above the mean-field level. This is related to
the one-body Wigner function which is obtained by a partial Fourier transform
acting on the relative coordinate $\mathbf{s}=\mathbf{r}-\mathbf{r}'$, i.e.
\begin{eqnarray}
 f^{(n)}_\mathrm{W}(\mathbf{r},\mathbf{k},t)
 &=&
  \int\frac{{\rm d}^n s}{(2\pi)^n}\:
  e^{-i\mathbf{k}\mathbf{s}} 
  \rho(\mathbf{r}\!-\!\frac{\mathbf{s}}{2},
       \mathbf{r}\!+\!\frac{\mathbf{s}}{2},t)
  \;,
\\
  \rho(\mathbf{r},\mathbf{r}',t)
  &=&
  \sum_{l}\Psi^\dagger_{l}(\mathbf{r},t)\Psi_{l}(\mathbf{r}',t)
\:.
\end{eqnarray}
Note that these are, in fact, a spin-averaged density matrix and correspondingly a
spin-averaged Wigner function. The dimensionality of the transformation is a
very compact notation and needs some explanation. Of course, our TDHF
calculations are always 3D. The full Wigner function is then a six-dimensional
object, obviously a bit bulky. Therefore, we often take cuts and look at the
Wigner transformation in reduced dimensions. The notation $f^{(1)}_\mathrm{W}$ then
means that one coordinate, e.g. $x$, is transformed from the pair $(x,x')$ in
the density matrix to the pair $(x,k_x)$ in the Wigner function. The other
two coordinates, $y$ and $z$ in the example, are fixed at a certain value $y_0$
and $z_0$, usually at the center of the nucleus $y_0=0$ and $z_0=0$. In other
words, $f^{(1)}_\mathrm{W}(x,k_x)$ denotes $\rho(x,y_0,z_0;x',y_0,z_0;t)$
transformed in the $x$ dimension.

A direct interpretation of the Wigner function as a phase-space probability
distribution is not possible because $f_\mathrm{W}$ is not positive
semidefinite.  There can arise situations where the quantum oscillations lead
to negative values.  These problems are avoided by the Husimi distribution
\cite{Takahashi,Toscano}. The Husimi function
$f_\mathrm{H}(\mathbf{r},\mathbf{k},t)$ is obtained by a convolution of the
Wigner function with a Gaussian $\mathcal{G}(\mathbf{r},\mathbf{k})$
\begin{eqnarray}
  && 
  f^{(n)}_\mathrm{H}(\mathbf{r},\mathbf{k},t) 
  = \int {\rm d}^n r'd^n k '\:
  \mathcal{G}(\mathbf{r}-\mathbf{r}',\mathbf{k}-\mathbf{k}')
\nonumber
\\
  &&\hspace*{6em}\times\:f^{(n)}_{W}(\mathbf{r}',\mathbf{k}',t)
  \:, 
\\
  &&
  \mathcal{G}^{(n)}
  (\mathbf{r}-\mathbf{r}',\mathbf{k}-\mathbf{k}')
  =
  \frac{1}{\pi^n}
  e^{-\frac{\mathbf{r}^2}{2\Delta{r}^2}}
  e^{-\frac{\mathbf{k}^2}{2\Delta{k}^2}}
  \:,
\\
  &&
  \Delta r\Delta k
  =
  \frac{1}{2}
  \:.
\end{eqnarray}
The Gaussian folding averages $f_\mathrm{W}$ over the minimal phase-space cell
of volume $(2\pi\hbar)^n$ and so successfully wipes out the negative
values. On the other hand, it induces some uncertainty which, however, is
physical because one cannot localize a particle in phase space better than
within a volume of $(2\pi\hbar)^n$.  The Husimi folding has one free
parameter, the folding width.  For best resolution in both directions it
should be chosen close to the width of the wave functions. As a basis for our
choice, we use here the nuclear harmonic oscillator model with frequency and
width parameter given as
\begin{equation*}
\hbar\omega
=
\frac{41\:\mathrm{MeV}}{A^{1/3}},
\quad 
\lambda
=
\frac{m\omega}{\hbar}.
\end{equation*}
This yields the estimate
\begin{eqnarray} 
  \Delta r^2
  &=&
  \frac{1}{2\lambda}
  =
  \frac{\hbar^2}{2m}{\hbar\omega}
  =
  \frac{A^{1/3}\:\mathrm{fm}^2}{2}
  \:,
\\
  \Delta k^2 
  &=&
  \frac{1}{4\Delta r^2}
  =
  \frac{\lambda}{2}
  =
  \frac{1}{2A^{1/3}\:\mathrm{fm}^2}
  \:.
\end{eqnarray}
The choice is somewhat ambiguous for nuclear reactions because one could
insert the mass number $A$ for the compound system or the average $A$ of
projectiles, or fragments respectively.  However, these are details which do
not hamper the analysis;  a good order-of-magnitude guess suffices for the present 
analysis.

\begin{figure}
\includegraphics*[width=8.3cm]{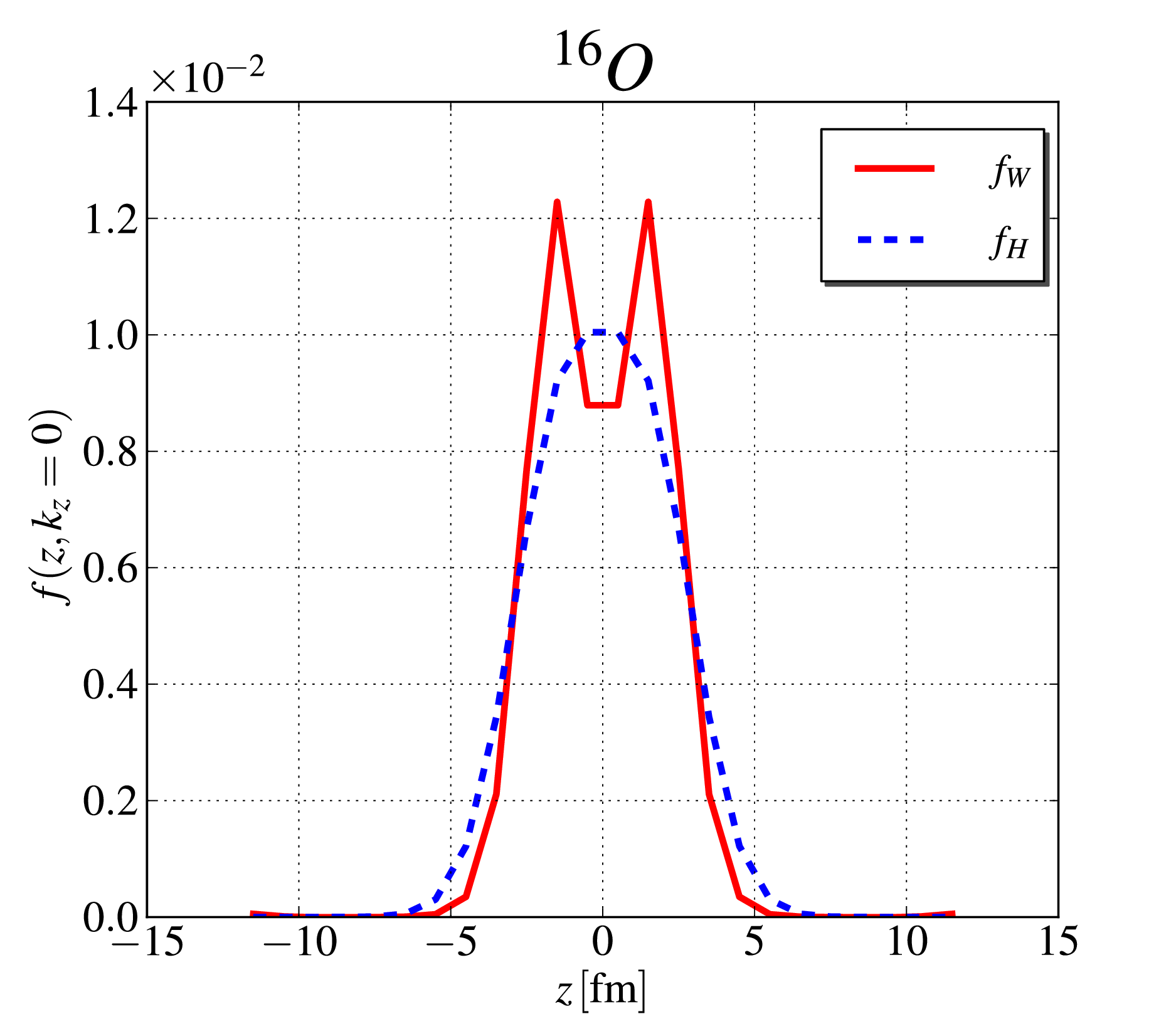}
\includegraphics*[width=8.3cm]{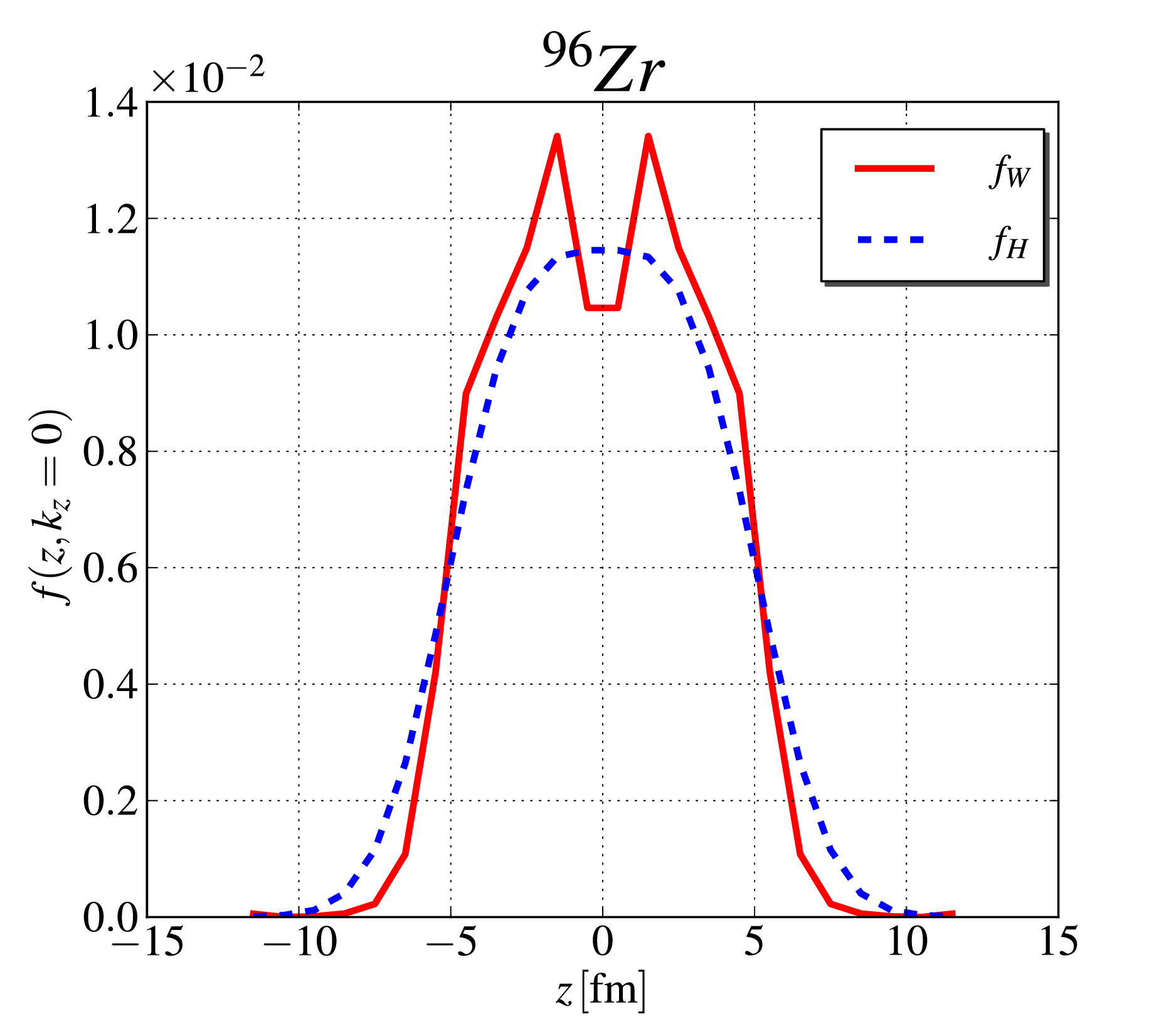}
\includegraphics*[width=8.3cm]{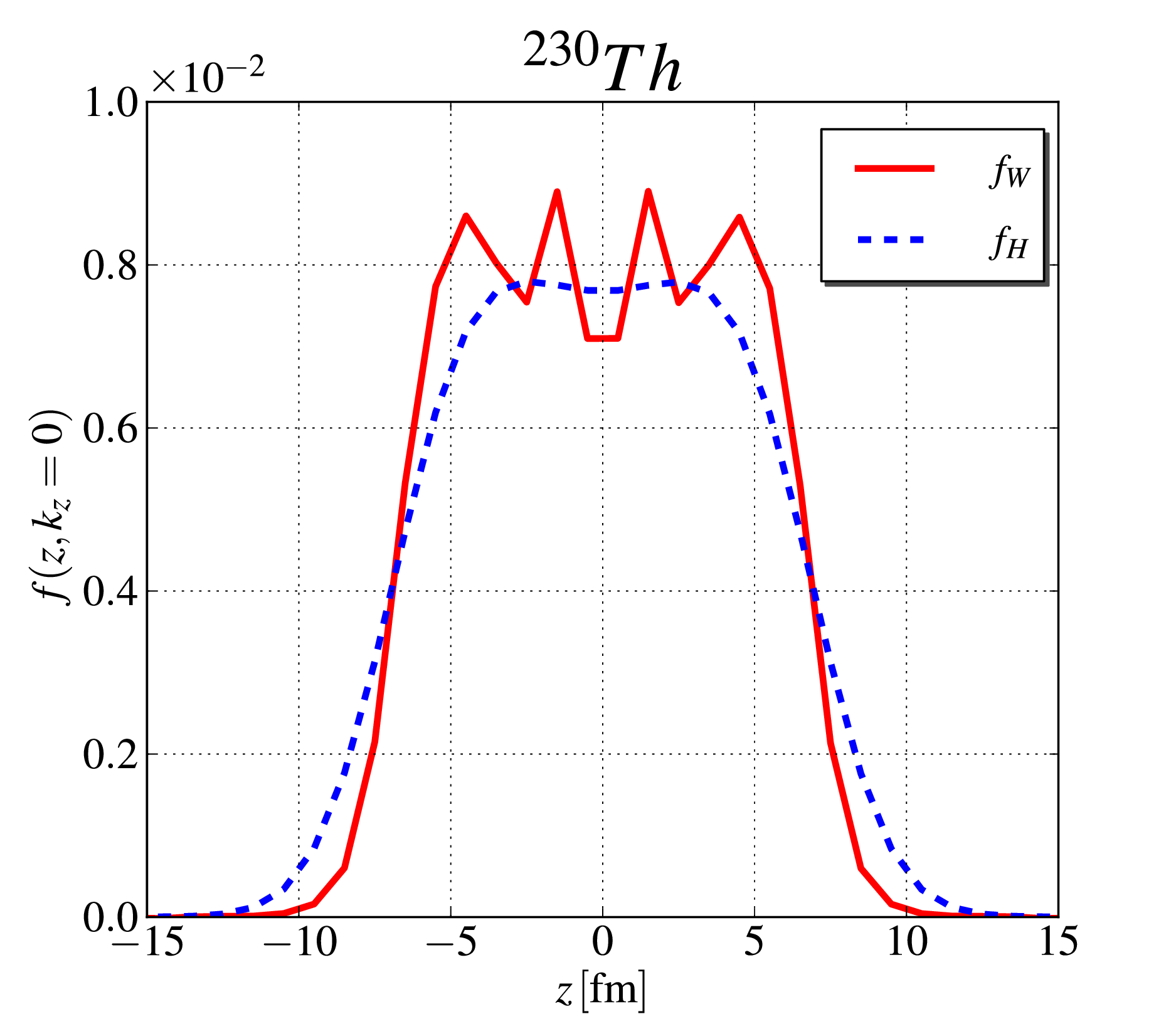}
\caption{\label{fig:static} (color online)
Comparison between slices through the one-dimensional
Wigner $f^{(1)}_\mathrm{W}(z,k_z=0)$ and Husimi $f^{(1)}_H(z,k_z=0)$
distribution functions for the static ground states of $^{16}O$ (top),
$^{96}Zr$ (middle), and $^{230}Th$ (bottom).}
\end{figure}

In a first round, we investigate the distributions for nuclear ground states,
to understand the basic pattern and to have a benchmark from a case certainly
free of excitation.  Figure \ref{fig:static} shows slices through static
one-dimensional Wigner and Husimi distributions of the ground states for three
nuclei, a light, a medium heavy, and a heavy one.  The Wigner distributions
show marked shell oscillations. The Husimi distributions have efficiently
removed these oscillations and represent a smooth curve averaged through the
Wigner distributions. The amplitude of the shell oscillations decreases with
increasing mass number, but very slowly such that smooth Wigner functions
(resembling classical phase-space distributions) are only reached at an order of
magnitude $A\approx 5000$ \cite{Bra97aB,Kri80a}.  The Husimi distributions
look smooth already for the low mass numbers of really existing nuclei. This,
however, is achieved at the price of somewhat blurring the details due to the
folding procedure.  This is acceptable for the analysis of the distributions
as such, i.e.  in phase space. It may become misleading when reducing the
distributions to compact observables by integrating over phase space or parts
thereof, as will be done in Section \ref{sec:observ}). The Husimi folding may
add an offset to such averaged observables. In such a case, the integrations
suffice to average out the small-scale oscillations in the Wigner functions.
Therefore, we will in the following concentrate our investigations on the 
use of the Wigner function only.

\section{More compact observables}
\label{sec:observ}

The Wigner and Husimi distributions are illustrative but difficult to handle, 
being six-dimensional objects. They can be
looked at in some selected snapshots and by taking cuts through the 6D phase
space. Observables in lower dimensions down to single numbers are necessary
complements for the analysis of dynamical processes. In this section, we will
introduce local observables which are distributed in 3D coordinate space.
They are reduced to single-number observables by further spatial integration.

\subsection{Observables from the Wigner distribution}

It is a standard procedure in classical non-equilibrium statistical physics to
discuss dissipation dynamics in terms of the local momentum distribution, i.e.
the momentum distribution at a given space point \cite{Bal75}. The basic
features of the local momentum distribution can be characterized by
its moments. The first moment  
\begin{equation}
  \langle\mathbf{k}(\mathbf{r},t)\rangle_{(n)}
  = 
  \frac{\int {\rm d}^n k\:\mathbf{k}\:f^{(n)}_\mathrm{W}(\mathbf{r},\mathbf{k},t)}
       {\int {\rm d}^n k\:f^{(n)}_\mathrm{W}(\mathbf{r},\mathbf{k},t)}.
\end{equation}
plays a special role. It characterizes the center of the distribution and it
is associated with the average local flow.  The higher moments are taken as
variances, i.e. relative to the first moment. For the $m$-th moment this reads
\begin{equation}
  \langle\mathbf{k}^{(m)}(\mathbf{r},t)\rangle_{(n)} 
  = 
  \frac{\int {\rm d}^n k \:(\mathbf{k}-\langle\mathbf{k}(\mathbf{r},
          t)\rangle)^{m}f^{(n)}_\mathrm{W}(\mathbf{r},\mathbf{k},t)}
       {\int {\rm d}^n k\:f^{(n)}_\mathrm{W}(\mathbf{r},\mathbf{k},t)}
  \:.
\end{equation}
These moments serve as raw material for further reduced observables.
Note that they depend on the dimensionality of the Wigner function
used in their definition. This is communicated by the index $(n)$ in
the moments.

The radial profile of the momentum distribution may be characterized
by the ratios of moments, in particular the $m=4$ to $m=2$ ratio
\begin{equation}
  R^{(n)}(\mathbf{r},t) 
  =
  \frac{\langle\mathbf{k}^{4}(\mathbf{r},t)\rangle_{(n)}}
       {\langle\mathbf{k}^{2}(\mathbf{r},t)\rangle^2_{(n)}}
  \:.
\label{eq:ratio}
\end{equation}
Reference value is the thermal equilibrium which corresponds in the high
temperature limit to a Maxwellian momentum distribution. These ``equilibrium''
values are given in Table \ref{tab:gauss} for various dimensions.

\begin{table}[htbp]
\caption{Analytic values of the ratio $R^{(n)}$ as defined in Eq.~(\ref{eq:ratio}) for
a Gaussian distribution function, depending on the spatial dimension
$n$ in which the ratio is evaluated.}
\begin{center}
\begin{tabular}{|c|c|}
\hline
dimension $n$ & $R^{(n)}_{\rm gauss}$ \\
\hline
$1$& $3$\\
$2$& $2$ \\
$3$& $5/3$ \\
\hline
\end{tabular}
\label{tab:gauss}
\end{center}
\end{table}

The ratios are plagued by the fact that cold equilibrium distributions are
Fermi functions rather than Gaussians and, more importantly, are significantly
smoothed by quantum effects. This hampers an analysis at a detailed level.  A
more robust signature of equilibration is obtained by the deformation of the
momentum distribution. The leading term is the quadrupole deformation which
can be characterized by the eccentricity in the reaction plane, which reads
\begin{equation}
   \varepsilon(\mathbf{r},t) 
  = 
  \frac{\langle k_x^{2}(\mathbf{r},t)\rangle-
          \langle k_z^{2}(\mathbf{r},t)\rangle}
       {\langle k_x^{2}(\mathbf{r},t)\rangle+
          \langle k_z^{2}(\mathbf{r},t)\rangle}\:,
\end{equation}
where the dimensionality index has been skipped for simplicity.
The global eccentricity is obtained by spatial integration
\begin{eqnarray}
  \varepsilon(t) 
  &=&  
  \int {\rm d}x\,dz\,\varepsilon(\mathbf{r},t)\rho(\mathbf{r},t)
  \:,
\\
 \rho(\mathbf{r})
 &=&
 \int {\rm d}k_x {\rm d}k_z\:f^{(2)}_\mathrm{W}(\mathbf{r},\mathbf{k},t)
 \:,
\end{eqnarray}
with $\rho$ the local density.

\subsection{Intrinsic kinetic energy}

Another interesting observable is the intrinsic excitation energy.  Ideally,
it is defined as the difference between the actual energy and a ``cold''
reference energy which is obtained from a stationary HF calculation
constrained to reproduce the density $\rho(\mathbf{r})$ and current
$\mathbf{j}(\mathbf{r})$ of the actual TDHF state \cite{Cus85a,Umar}. The
cumbersome density constrained calculations can be avoided when evaluating the
``cold'' reference state in Thomas-Fermi approximation. This shortcut was used
successfully in Cluster physics \cite{Calvayrac}. The so approximated
intrinsic kinetic energy reads
\begin{eqnarray}
  E_{\rm int}(t)
  &=&
  E_{\rm kin}(t)-E_{\rm coll,kin}(t)-E_{\rm TFW}(t)
  \:,
\\
  E_{\rm kin}(t)
  &=&
 \frac{1}{2}\sum_i \int {\rm d}^3r\:|\nabla \varphi_i({\bf r},t)|^2
 \:, 
\\
  E_{\rm coll,kin}(t)
  &=&
  \int {\rm d}^3 r \: \frac{{\bf j}^2({\bf r},t)}{2 \rho({\bf r},t)}
  \:,
\\
  E_{\rm TFW}(t)
  &=&
  \int {\rm d}^3r\: \tau_{\rm TFW}({\bf r},t)
  \;, 
\\
  \tau_{\rm TFW}({\bf r},t) 
  &=& 
  \frac{3\hbar^2}{10m}(3\pi^2)^{2/3} \rho({\bf r},t)^{5/3} 
\nonumber \\
  && 
 +\frac{\hbar^2}{18m}\frac{(\nabla\rho({\bf r},t))^2}{\rho({\bf r},t)}
 \;.
\end{eqnarray}
It quantifies the non-adiabatic and non-collective component of the kinetic
energy, roughly corresponding to the intrinsic thermal energy.  The first
ingredient for the calculation is the total kinetic energy, $E_{\rm kin}$, of
the system.  The second term, $E_{\rm coll,kin}(t)$, subtracts the 
hydrodynamic kinetic energy contained in the collective flow $\mathbf{j}$.
The third term, $E_{\rm TFW}(t)$, subtracts the instantaneous kinetic energy
of the zero-temperature ground state at the given density $\rho({\bf
  r},t)$. The evaluation of this kinetic energy density $\tau({\bf r},t)$ is
done in the Thomas-Fermi-Weizs\"acker approximation \cite{Dreizler}.

\begin{figure*}[hbtp]
\includegraphics*[width=8.6cm]{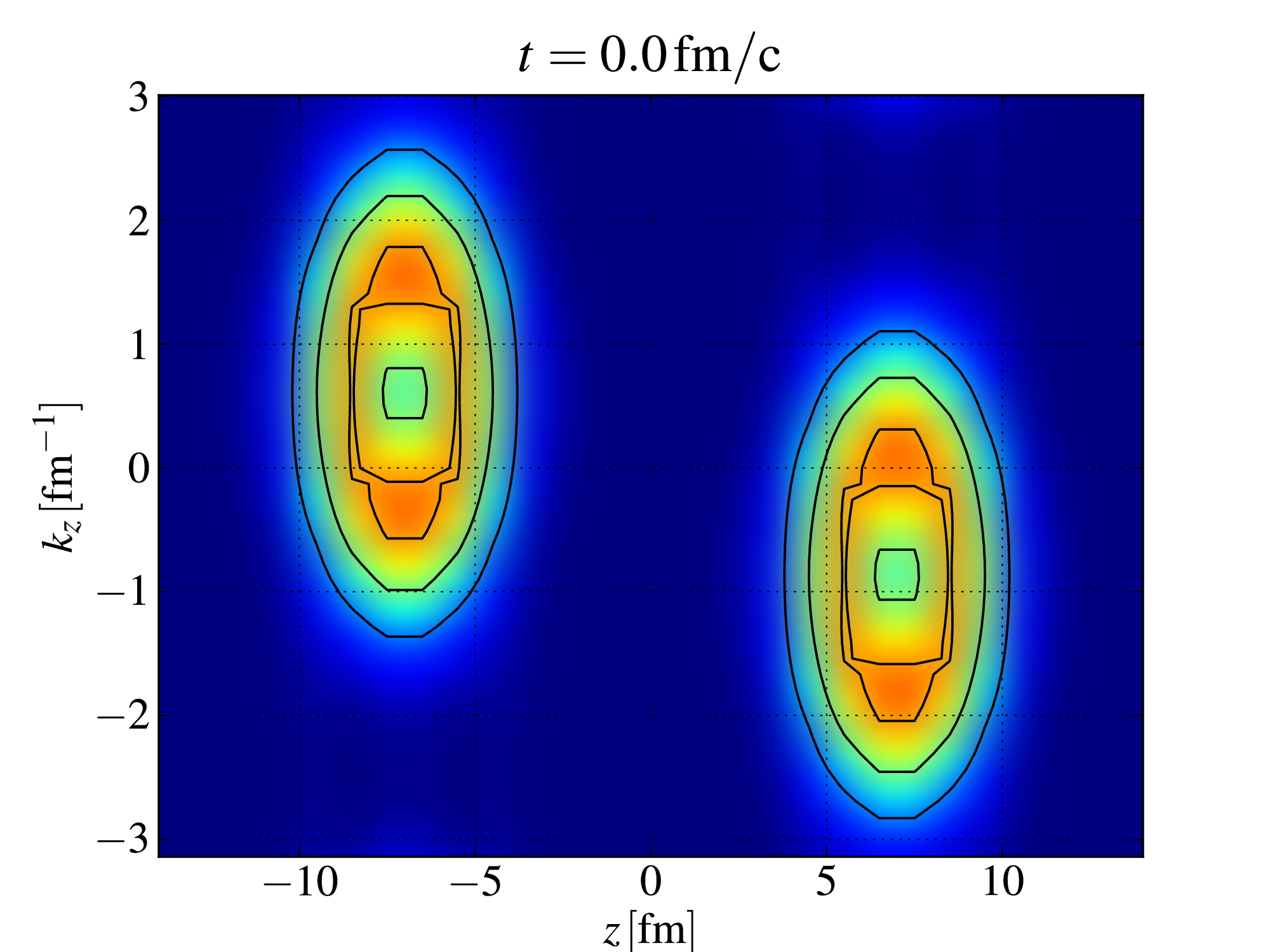}
\includegraphics*[width=8.6cm]{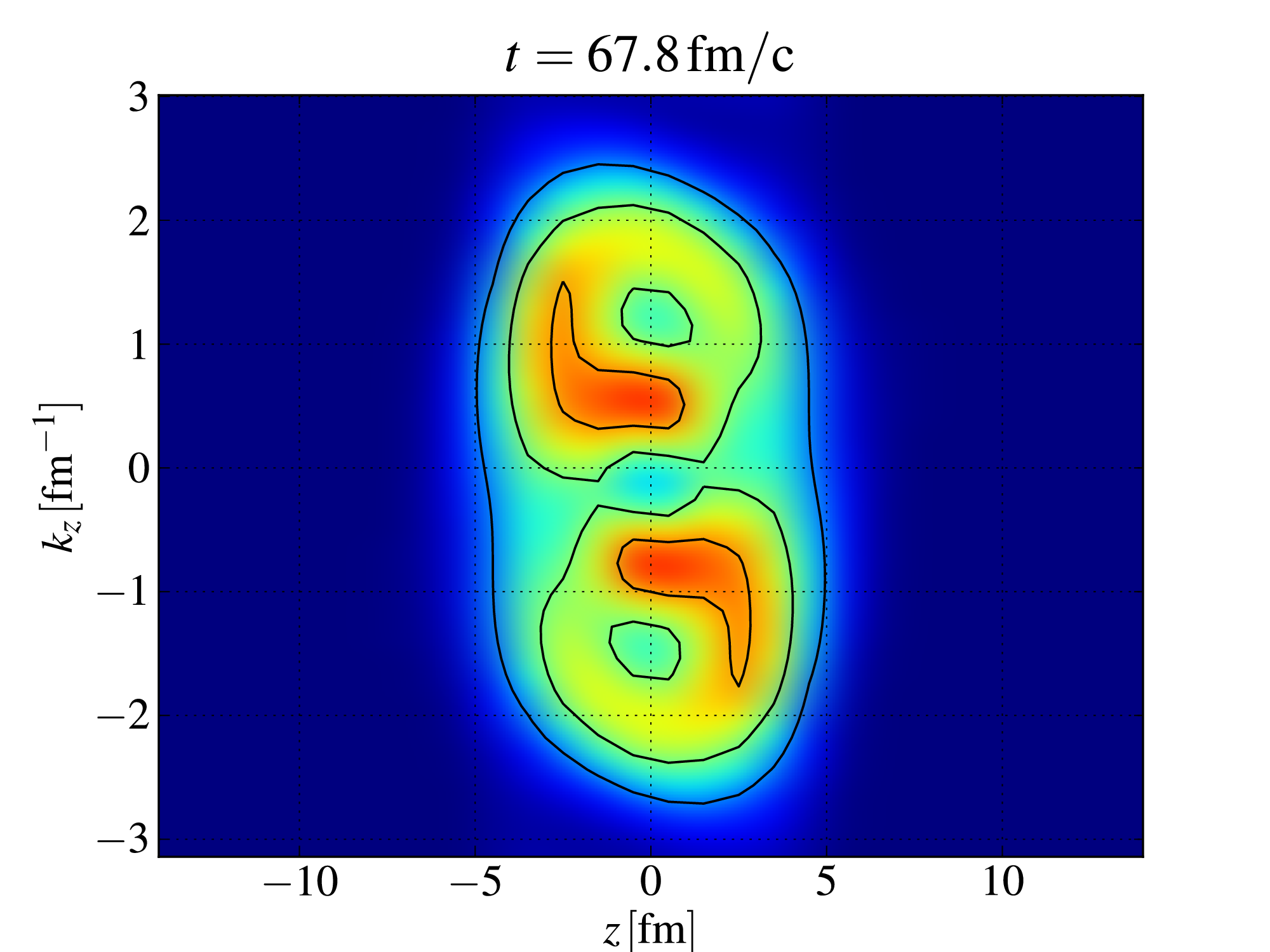}
\includegraphics*[width=8.6cm]{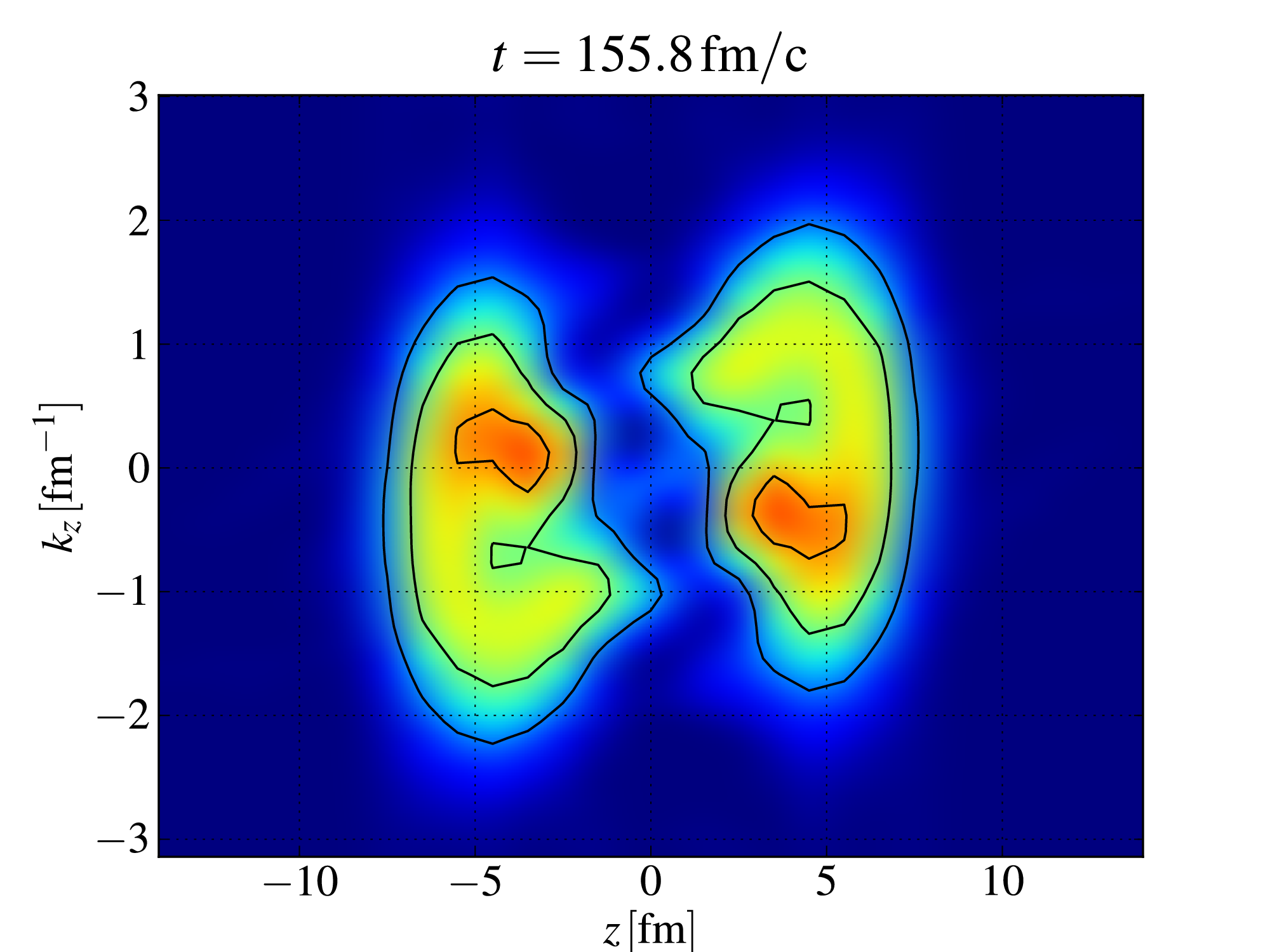}
\includegraphics*[width=8.6cm]{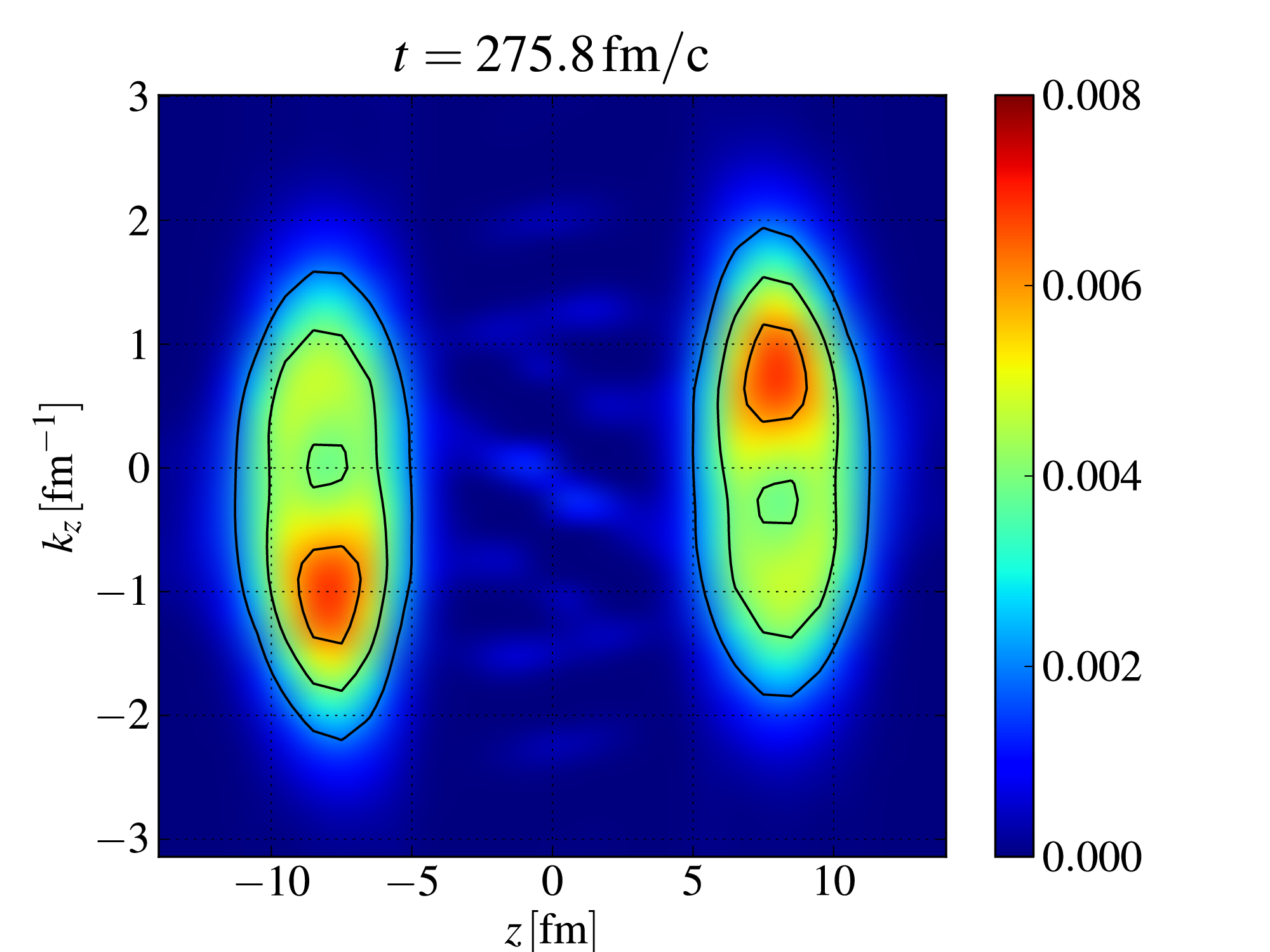}
\caption{\label{fig:O16} (color online)
The one-dimensional Wigner distribution
$f^{(1)}_\mathrm{W}(z,k_z,t)$ for a central $^{16}O+$$^{16}O$ collision
is plotted at four different times $t$. Three contour lines are
plotted to highlight the levels of $f^{(1)}_\mathrm{W}(z,k_z,t)$ at 
$2\cdot10^{-3},\:4\cdot10^{-3}$,\: and $6\cdot10^{-3}$.}
\end{figure*}

\subsection{An estimate for the fragment distance}

As a simple observable characterizing the geometry of a collisional stage, we
introduce the distance $d(t)$ between the fragments
\begin{equation}
  d(t) 
  = 
  |\langle\mathbf{r}_1(t)\rangle-\langle\mathbf{r}_2(t)\rangle|
  \:.
\end{equation}
The coordinates $\mathbf{r}_i $ of the right and left fragment were obtained
by splitting the density of the system symmetrically into two half spaces and
averaging over each half. This is an obvious definition for well separated
fragments. It becomes somewhat ambiguous in the overlap region, but still
remains a useful indicator of the overall geometry.

\section{Results}
\label{sec:results}

We present TDHF results for different reaction scenarios, $^{16}O+$$^{16}O$
collisions head-on and with finite impact parameter, and a
$^{96}Zr+$$^{132}Sn$ collision. The Skyrme parametrization SkI3
\cite{Reinhard2} is used for the calculations.  We performed test calculations
with other Skyrme forces and found very similar results. Thus we report the
results only from this one force.  The central collisions were computed on a
coordinate space mesh with $48\times24^2$ grid points and the non-central ones
with $36\times24^2$ points.

\subsection{$^{16}$O+$^{16}$O Collisions}

\subsubsection{$^{16}O+$$^{16}O$ Central}

First, we analyze a $^{16}O+$$^{16}O$ collision with a center-of-mass energy
of $E_{c.m.}=100$\:MeV and zero impact parameter $b=0$\:fm. Figure
\ref{fig:O16} shows the one-dimensional Wigner distribution
$f^{(1)}_\mathrm{W}(z,k_z,t)$ at four different stages of the collision.
Initially ($t=0.0$\:fm/c), there are two cold nuclei far apart from each
other. They are shifted in $k_z$-direction depending on their initial
boost. At the intermediate stage ($t=67.8$\:fm/c), the phase space volumes of
the two fragments seem to merge but are avoiding each other, i.e. they
maintain a division line. This is a consequence of the Pauli principle. After
a while ($t=155.8$\:fm/c), the phase space volumes start to separate, keeping
some contact still for some time.

The final stage ($t=273.8$\:fm/c) shows two separate fragments again.  But
here the structure is quite different as compared to the initial state.  Both
blobs become strongly asymmetric.  The $k_z$ position of the maximal peaks
(red spots) are not lower than initially. But the asymmetry in the
distribution extends very much towards lower $k_z$ and also to values of
opposite sign. This indicates that the average slowdown in relative
c.m. motion in this case is not due to a global downshift of an otherwise
symmetric distribution, but to the strong asymmetry reducing significantly
the average $k_z$. A possible interpretation is that the
wave functions maintain their initial momentum structure to a large extent 
but also components from the other fragment are mixed in.

\begin{figure}[htbp]
\includegraphics*[width=8.6cm]{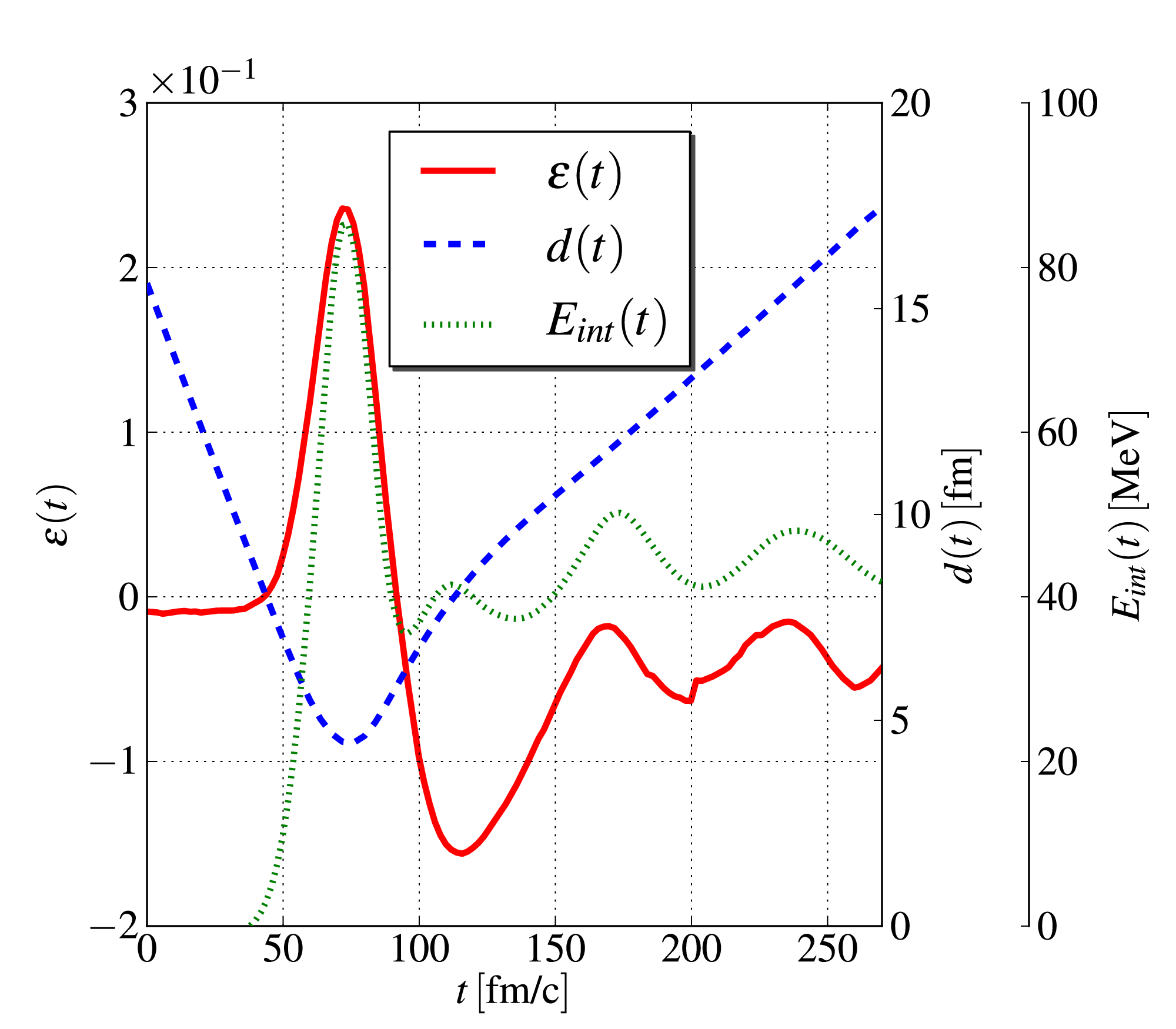}
\caption{\label{fig:eccent_eint} (color online)
The global eccentricity $\varepsilon(t)$ obtained from the
two-dimensional Wigner function $f^{(2)}_\mathrm{W}(x,z,k_x,k_z,t)$,  the
distance between the fragments $d(t)$, and the internal kinetic
energy $E_{int}(t)$  for a central $^{16}O+$$^{16}O$
collision with a center-of-mass energy of $E_{c.m.}=100$\:MeV.}
\end{figure}

In a next step, we analyze the time evolution in terms of compact (single
number) observables.  The time evolution of the intrinsic kinetic energy
$E_{int}(t)$ and of the global eccentricity $\varepsilon(t)$ are
shown, together with the fragment distance $d(t)$, in Figure
\ref{fig:eccent_eint}.  The time of maximum overlap (compound stage) is
reached at 75 fm/c where $d(t)$ has a minimum.  Both kinetic observables show
a pronounced maximum there.  As the reaction continues the eccentricity is
strongly damped and keeps oscillating at a low level.  This indicates some
thermalization.  The internal energy reaches its maximum at
$E_{int}\approx86$\:MeV and saturates, again with some persisting
oscillations, at half of the maximal amount. As the potential energy plays a
huge role in the compound stage, the values for the kinetic energies have to
be taken with care here. The asymptotic values are more directly
interpretable. They show significant heating (from $E_{int}(t)$) and
great deal of equilibration already within the short time span of the
simulation.   

\begin{figure}[htbp]
\includegraphics*[width=8.1cm]{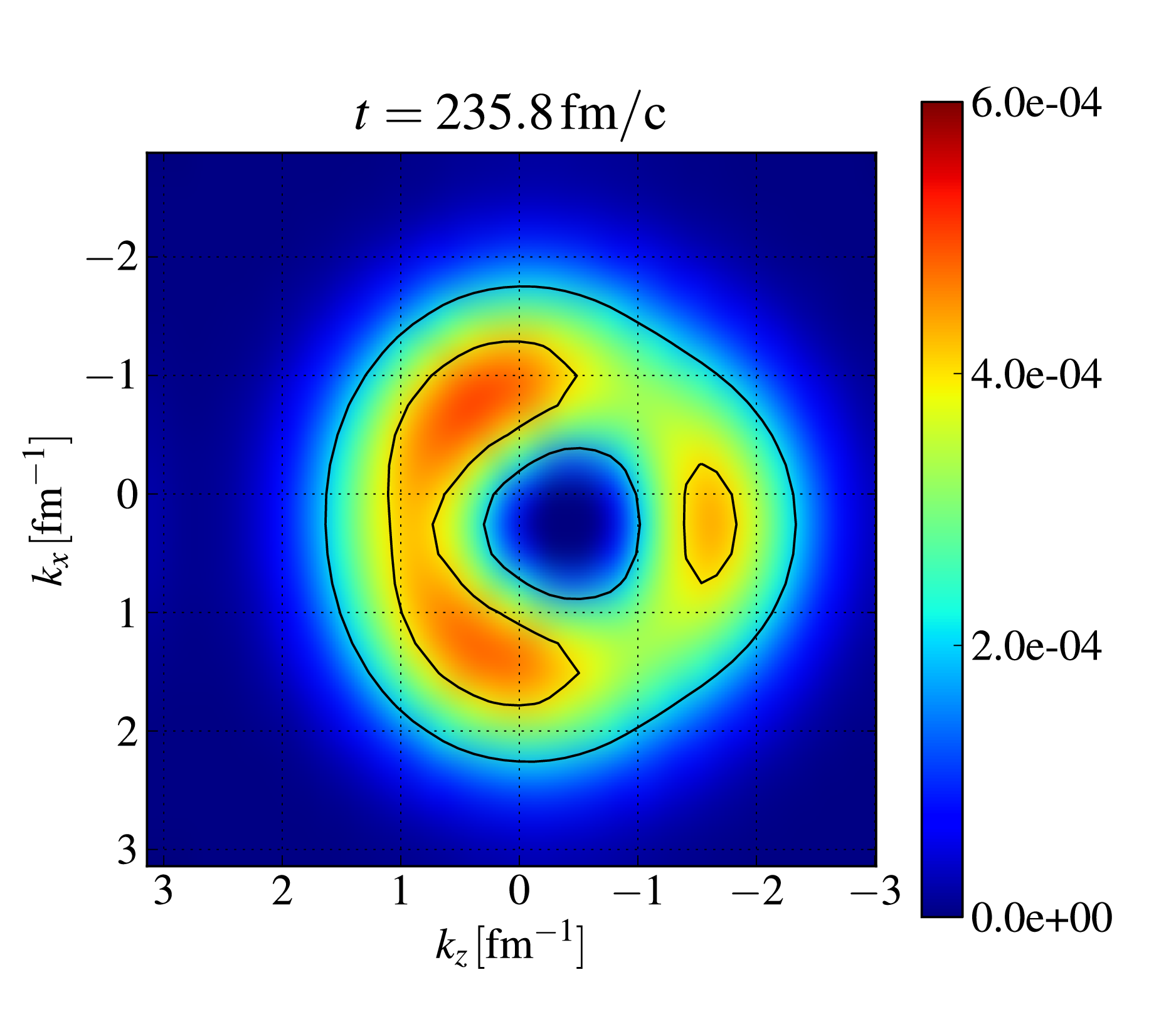}
\includegraphics*[width=8.1cm]{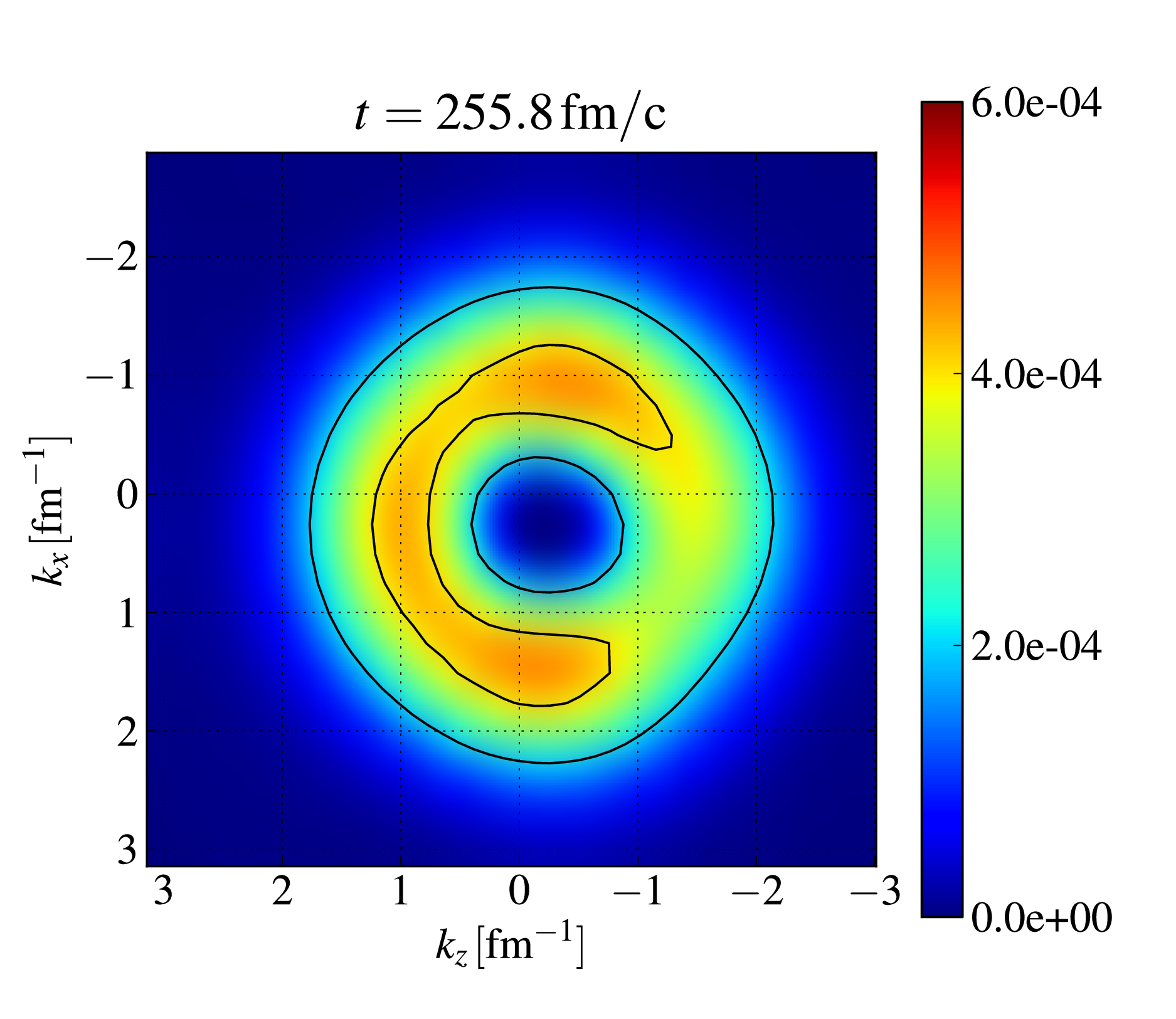}
\includegraphics*[width=8.1cm]{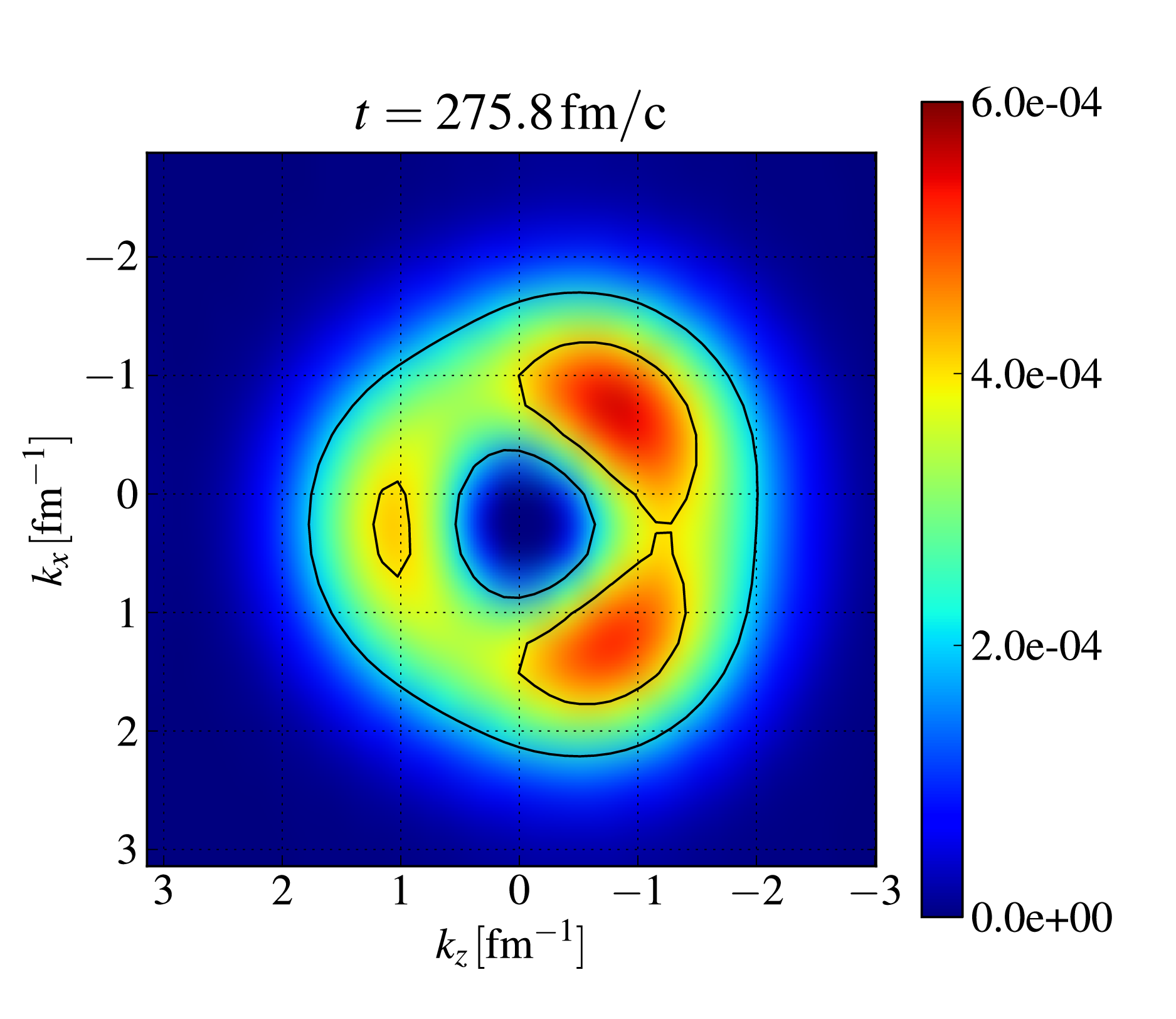}
\caption{\label{fig:Osci} (color online)
A $k_x$-$k_z$ cut of the
 two-dimensional momentum distribution $f^{(2)}_\mathrm{W}(k_x,k_z,t)$
plotted at the center of the fragment moving finally with
negative mean momentum in $k_z$-direction. The plots are taken at
three different times $t$ near the final stage of the calculation. Two
contour lines are plotted to highlight the levels of $f^{(2)}_\mathrm{W}$ at
$2\cdot10^{-4}$ and $4\cdot10^{-4}$.}
\end{figure}

To visualize the oscillations in $\varepsilon(t)$ and $E_{int}(t)$ we show in
Figure \ref{fig:Osci} the two-dimensional momentum distribution
$f^{(2)}_\mathrm{W}(k_x,k_z,t)$ in the exit channel for the fragment moving to
the left (with negative $\langle k_z\rangle$).  The shape is
asymmetric and oscillates back and forth.  This indicates that the largest
collective effect in the exit channel is residual octupole oscillations which
have their counterpart also in similar octupole oscillations of the fragments'
spatial shape.

We have checked the momentum ratio $R^{(n)}(\mathbf{r},t)$ as given in
Eq. (\ref{eq:ratio}) for $n=1,2$ at different times to probe the
closeness of the momentum distribution to a Maxwellian
distribution. The comparison with the analytic values $R_{\rm gauss}$ from
Tab. \ref{tab:gauss} is not conclusive, as quantum
effects blur the classical concept behind this ratio by making the distributions 
too different from Gaussians. This casts some
doubts on the usefulness of the global ratio $R(t)$ in this still
predominantly quantum-mechanical domain. We will come back to this
observable in Section \ref{sec:ZrSn}.

\subsubsection{$^{16}$O+$^{16}$O Non-Central}

In this Section we analyze $^{16}$O+$^{16}$O fusion reactions with
non-zero impact parameter $b=2$\:fm and two different center-of-mass
energies $E_{c.m.}=20$\:MeV and $E_{c.m.}=50$\:MeV. 

\begin{figure}[htbp]
\includegraphics*[width=8.6cm]{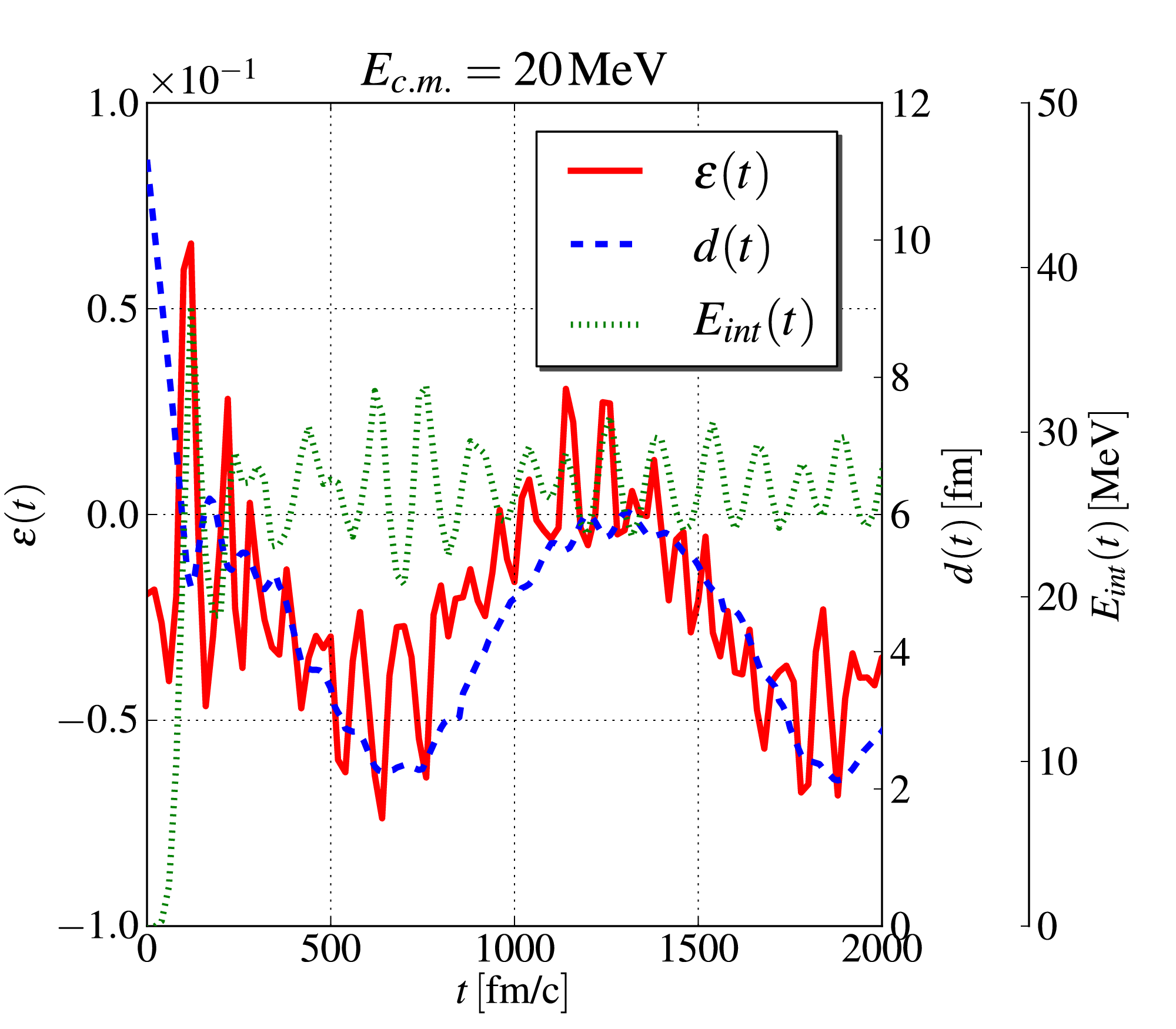}
\includegraphics*[width=8.6cm]{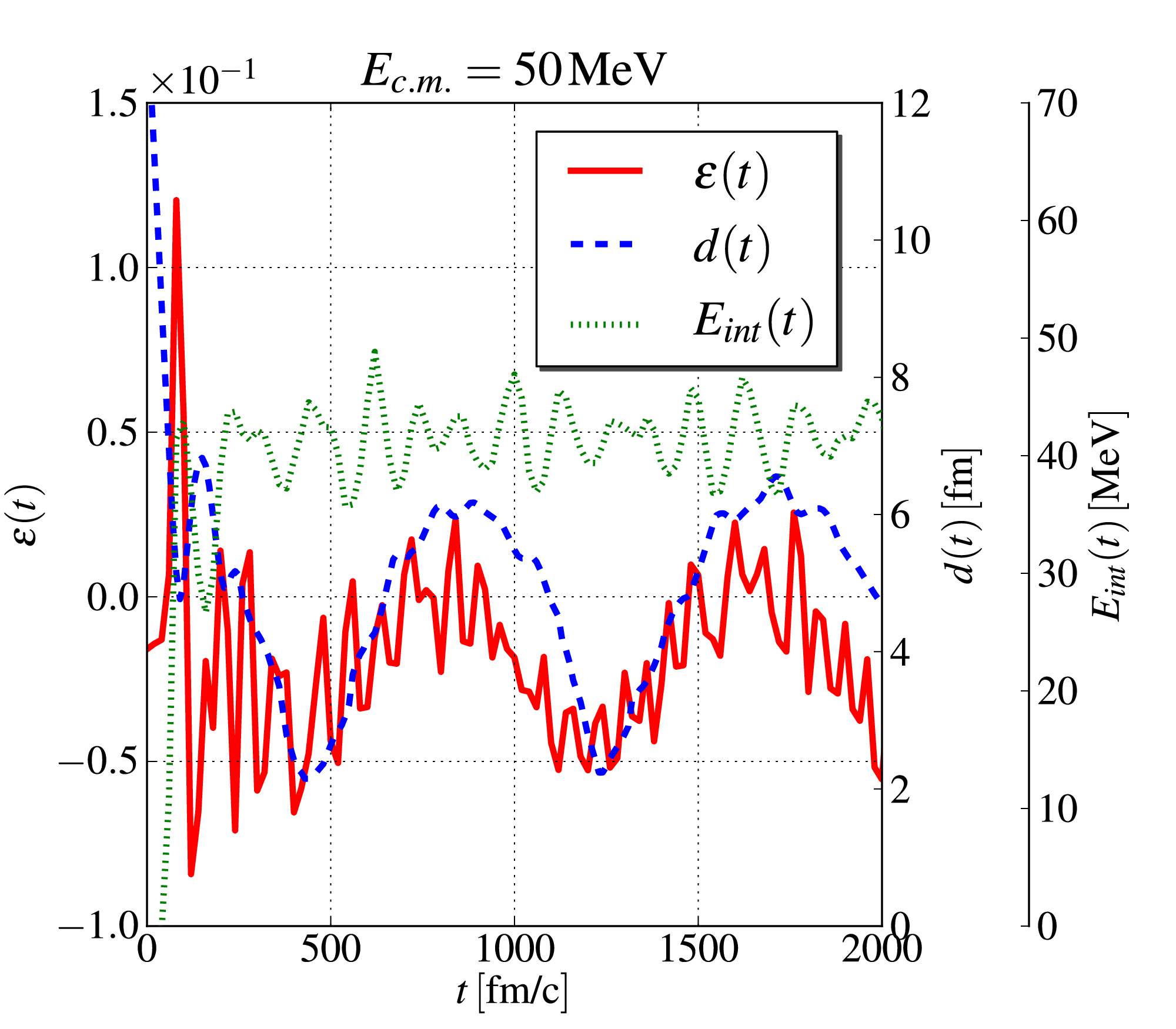}
\caption{\label{fig:fusion} (color online)
The global eccentricity $\varepsilon(t)$ obtained
from the two-dimensional Wigner function $f^{(2)}_\mathrm{W}(x,z,k_x,k_z,t)$, 
the distance between the fragments $d(t)$, and the internal
kinetic energy $E_{int}(t)$ for two $^{16}$O+$^{16}$O
fusion reaction with a center-of-mass energy of $E_{c.m.}=20$\:MeV
(top) and $E_{c.m.}=50$\:MeV (bottom) and impact parameter $b=2$\:fm.}
\end{figure}

\begin{figure}[htbp]
\includegraphics*[width=8.6cm]{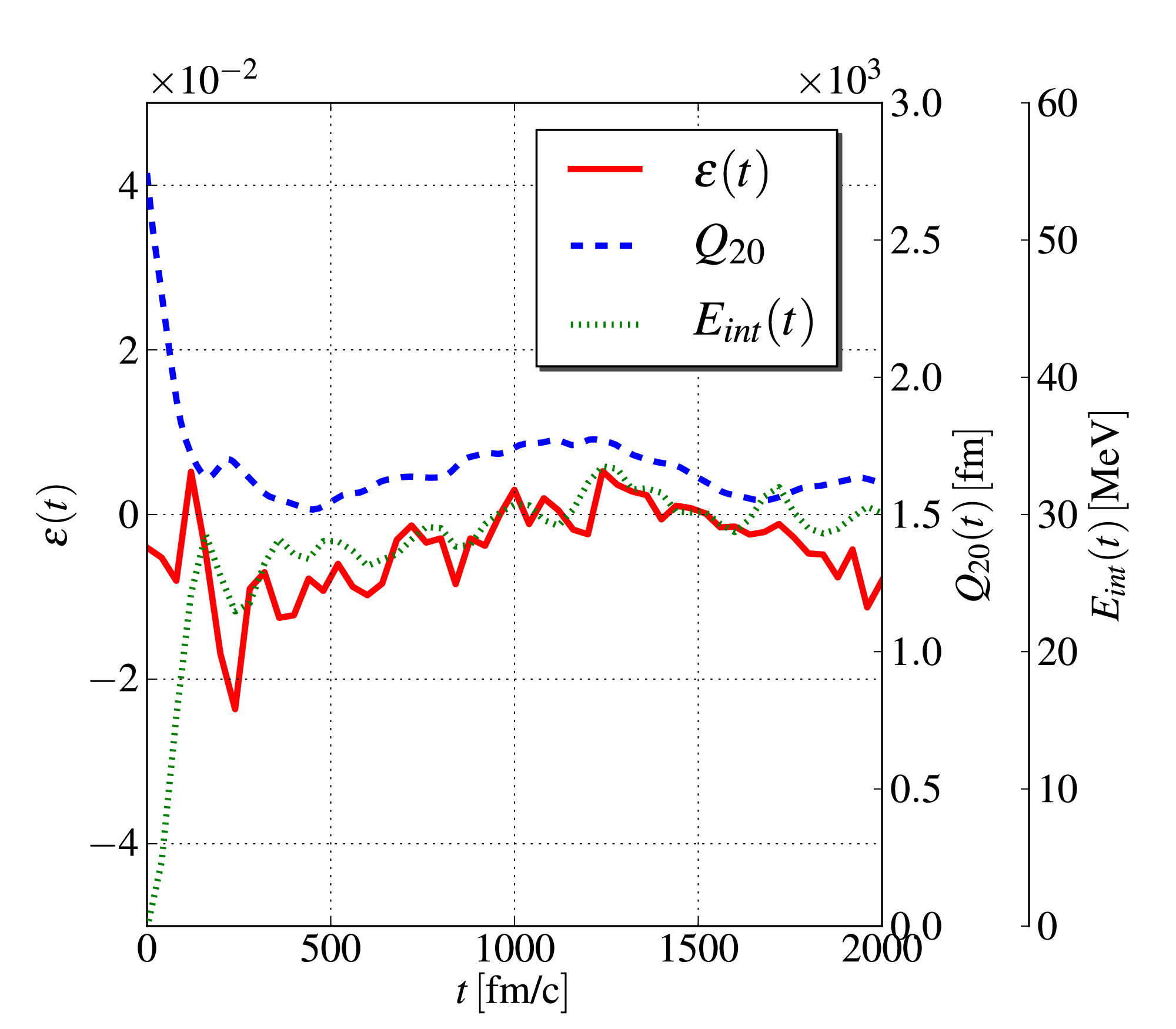}
\caption{\label{fig:ZrSn} (color online)
The global eccentricity $\varepsilon(t)$ obtained from the
two-dimensional Wigner function $f^{(2)}_\mathrm{W}(x,z,k_x,k_z,t)$, the
quadrupole $Q_{20}$, and the internal kinetic energy $E_{int}(t)$
for a $^{96}$Zr+$^{132}$Sn fusion reaction with a center-of-mass 
energy of $E_{c.m.}=250$\:MeV and impact parameter $b=2$\:fm.}
\end{figure}

\begin{figure}[htbp]
\includegraphics*[width=8.6cm]{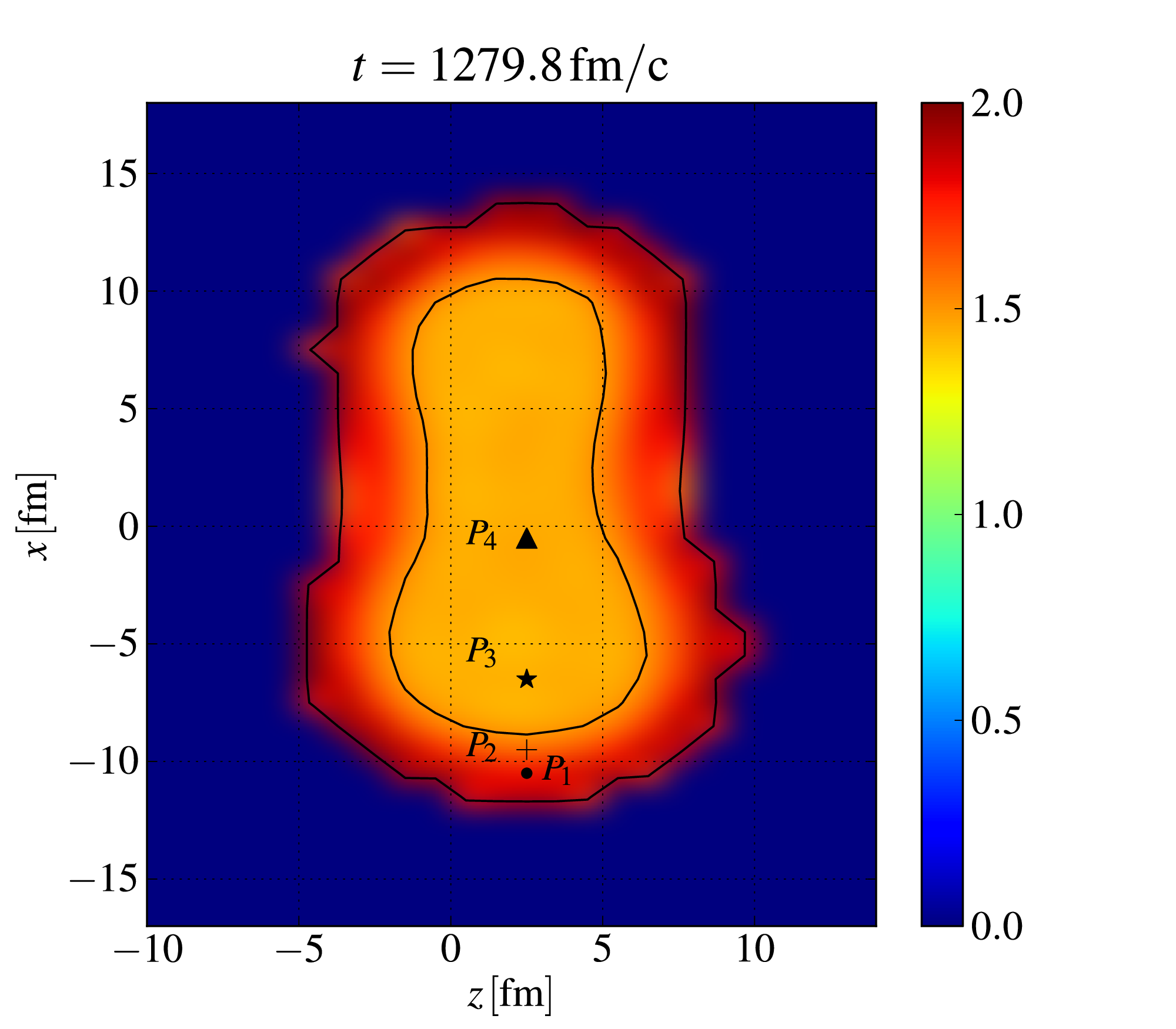}
\caption{\label{fig:ratio2d} (color online)
The local ratio $R^{(2)}(\mathbf{r})$ obtained from the
two-dimensional Wigner distribution $f^{(2)}_\mathrm{W}(x,z,k_x,k_z,t)$ for
a $^{96}Zr+$$^{132}Sn$ fusion reaction at
$t=1279.8$\:fm/c. Four points are marked in this plot to be analyzed
later more precisely (Figure \ref{fig:ZrSnMomentA} and Figure
\ref{fig:ZrSnMomentB}). The reference value from a Gaussian
distribution is $R^{(2)}_{\rm gauss}=2$. A Contour line is plotted to
highlight the level of $R^{(2)}(\mathbf{r})=1.5$.}
\end{figure}

\begin{figure*}[!htbp]
\includegraphics*[width=8.6cm]{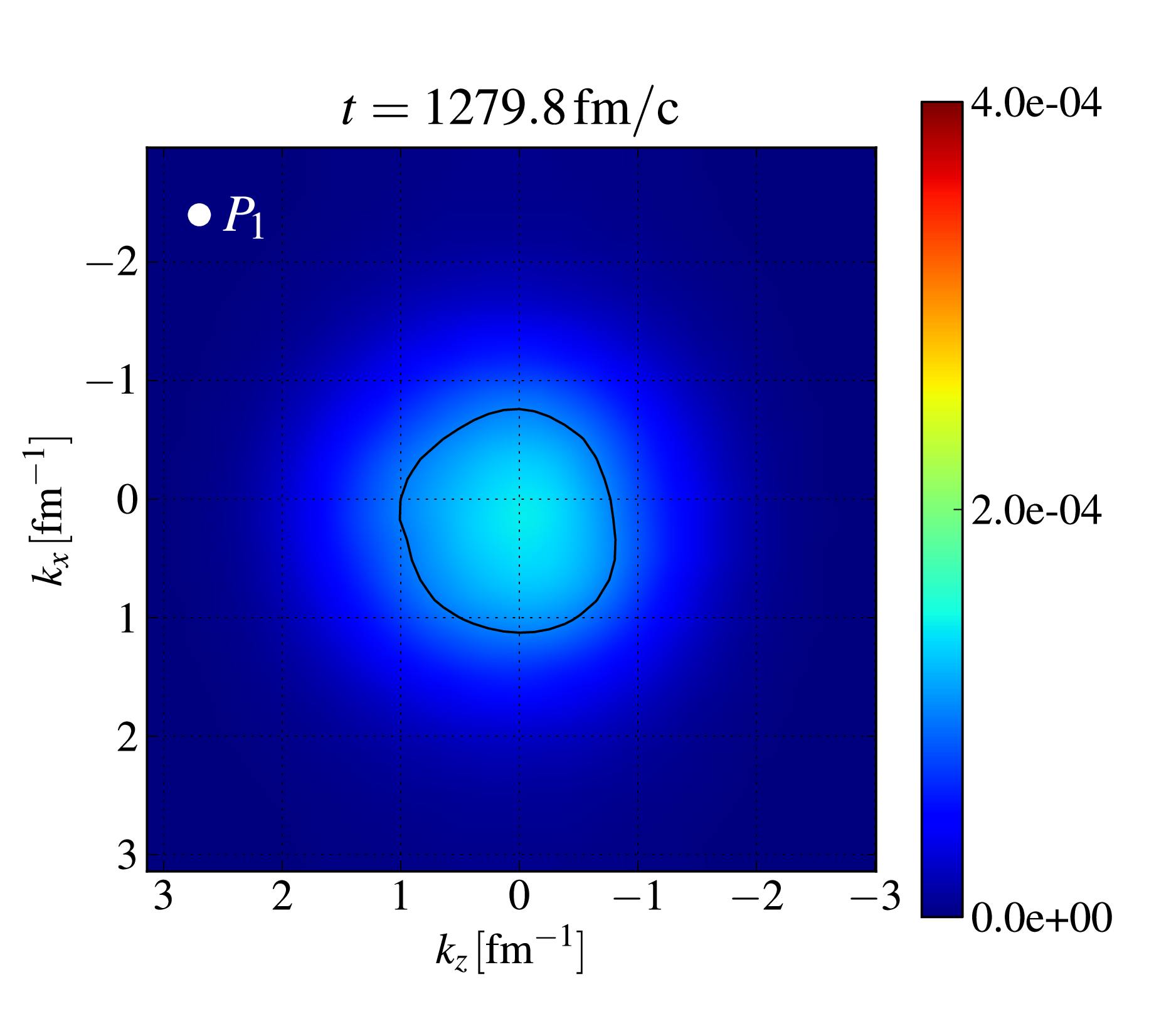}
\includegraphics*[width=8.6cm]{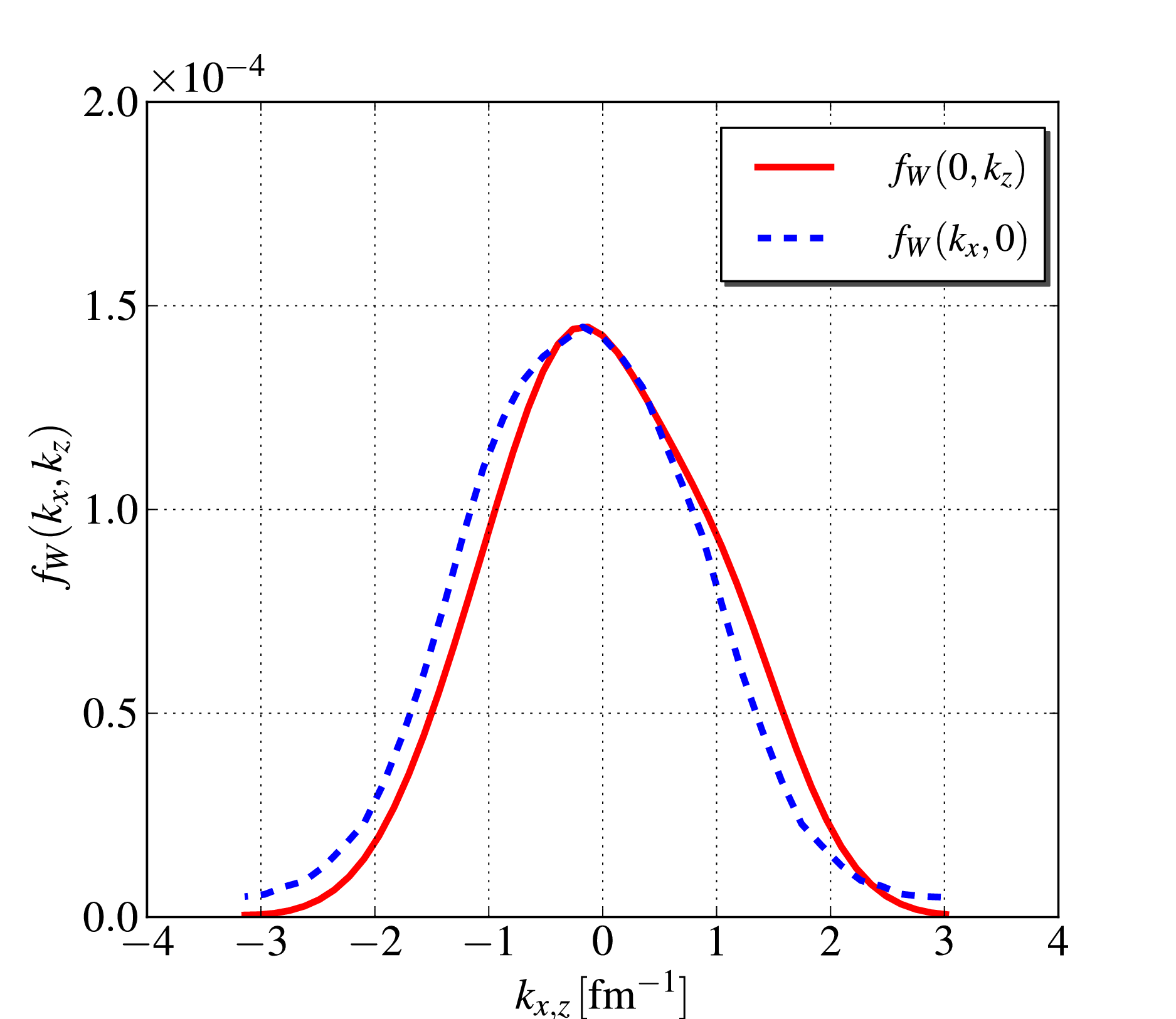}
\includegraphics*[width=8.6cm]{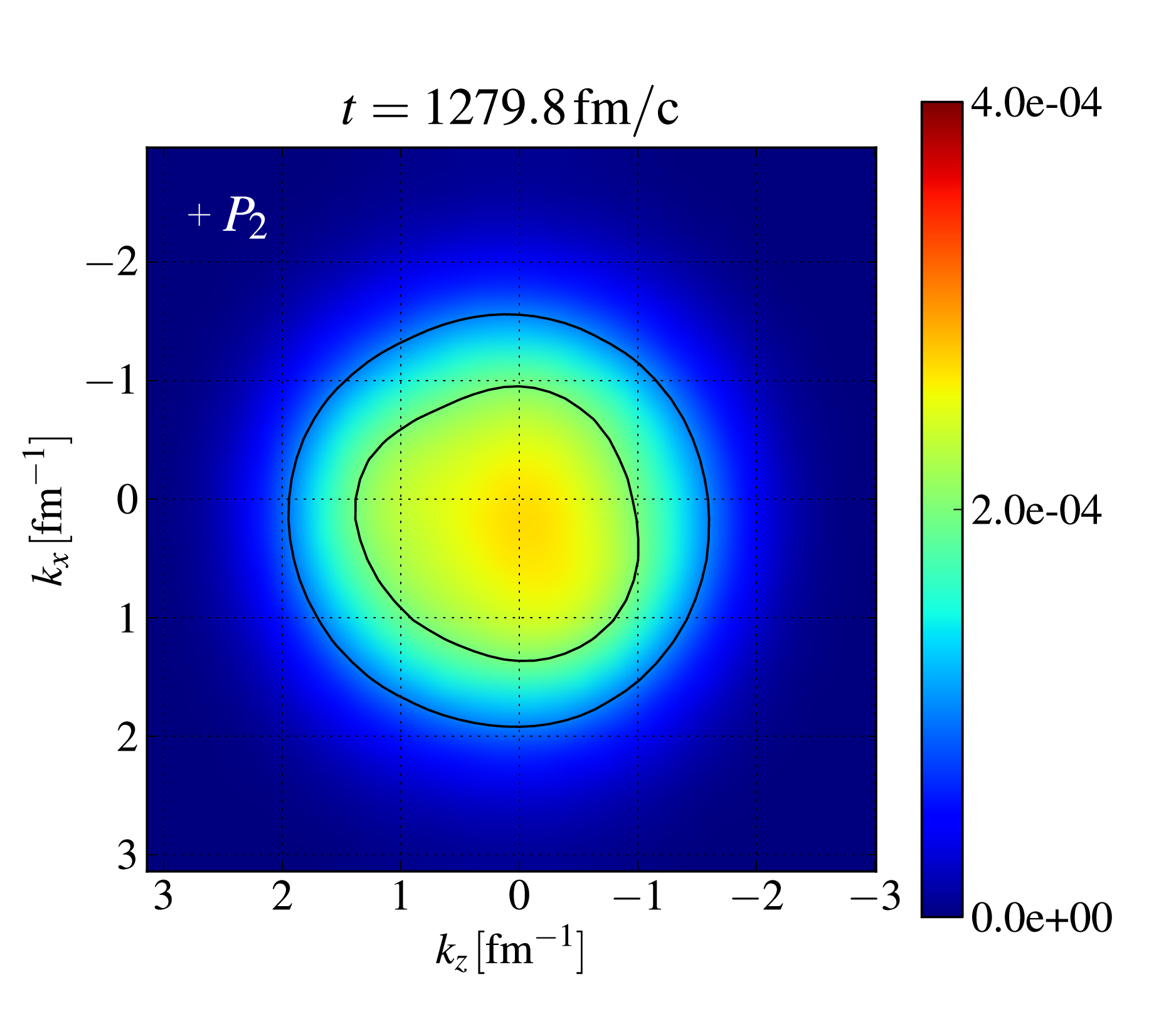}
\includegraphics*[width=8.6cm]{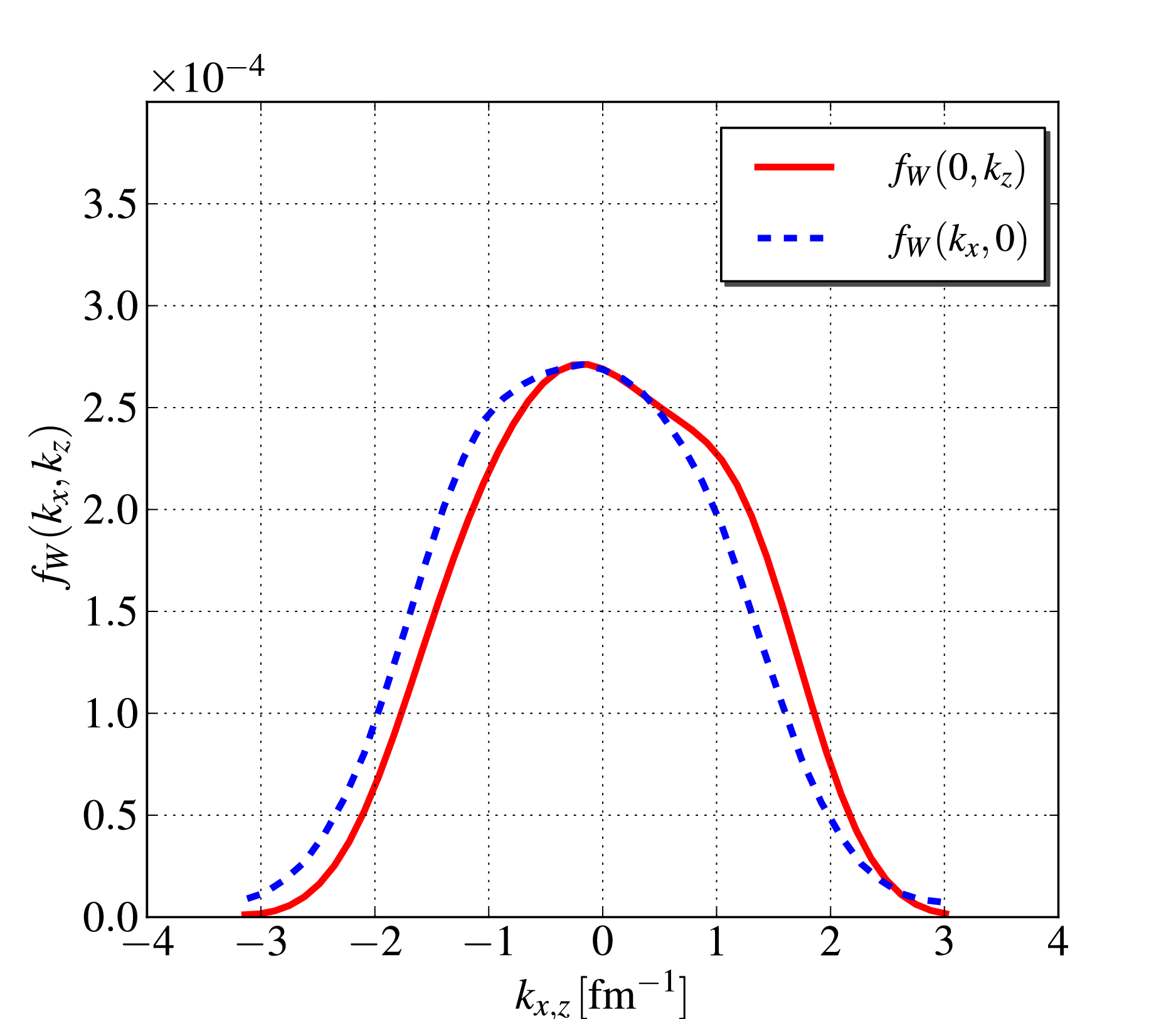}
\caption{\label{fig:ZrSnMomentA} (color online)
The left column reviews the two-dimensional momentum
distribution $f^{(2)}_\mathrm{W}(k_x,k_z)$ in the selected (outer) points from
Figure \ref{fig:ratio2d}. Contour lines are plotted to highlight the
levels of $f^{(2)}_\mathrm{W}$ at $1\cdot10^{-4},2\cdot10^{-4}$, and
$3\cdot10^{-4}$.  Slices through $f^{(2)}_\mathrm{W}(k_x,k_z)$ matching the
$k_x,k_z$-axis are shown in the right column.}
\end{figure*}

\begin{figure*}[!htbp]
\includegraphics*[width=8.6cm]{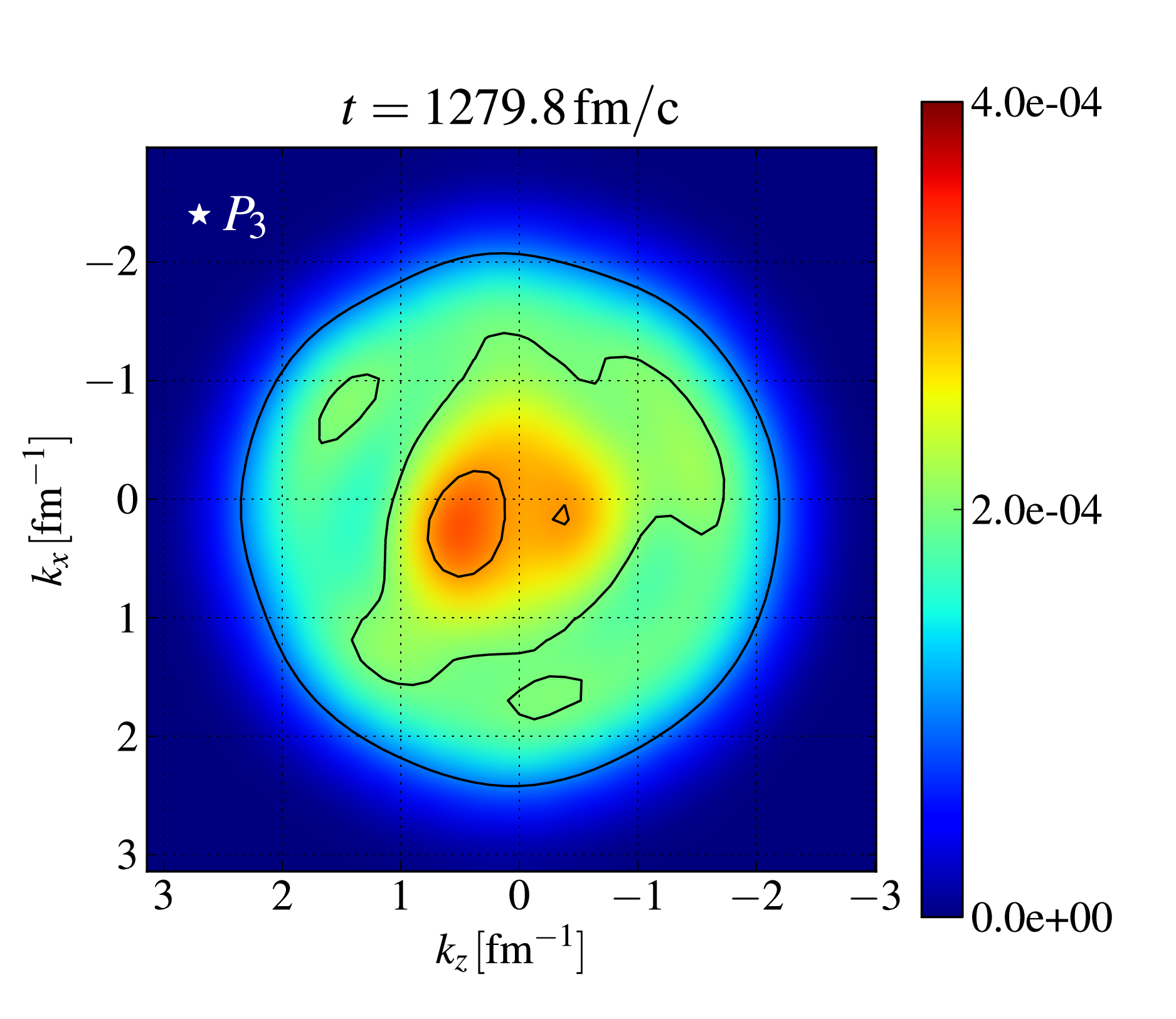}
\includegraphics*[width=8.6cm]{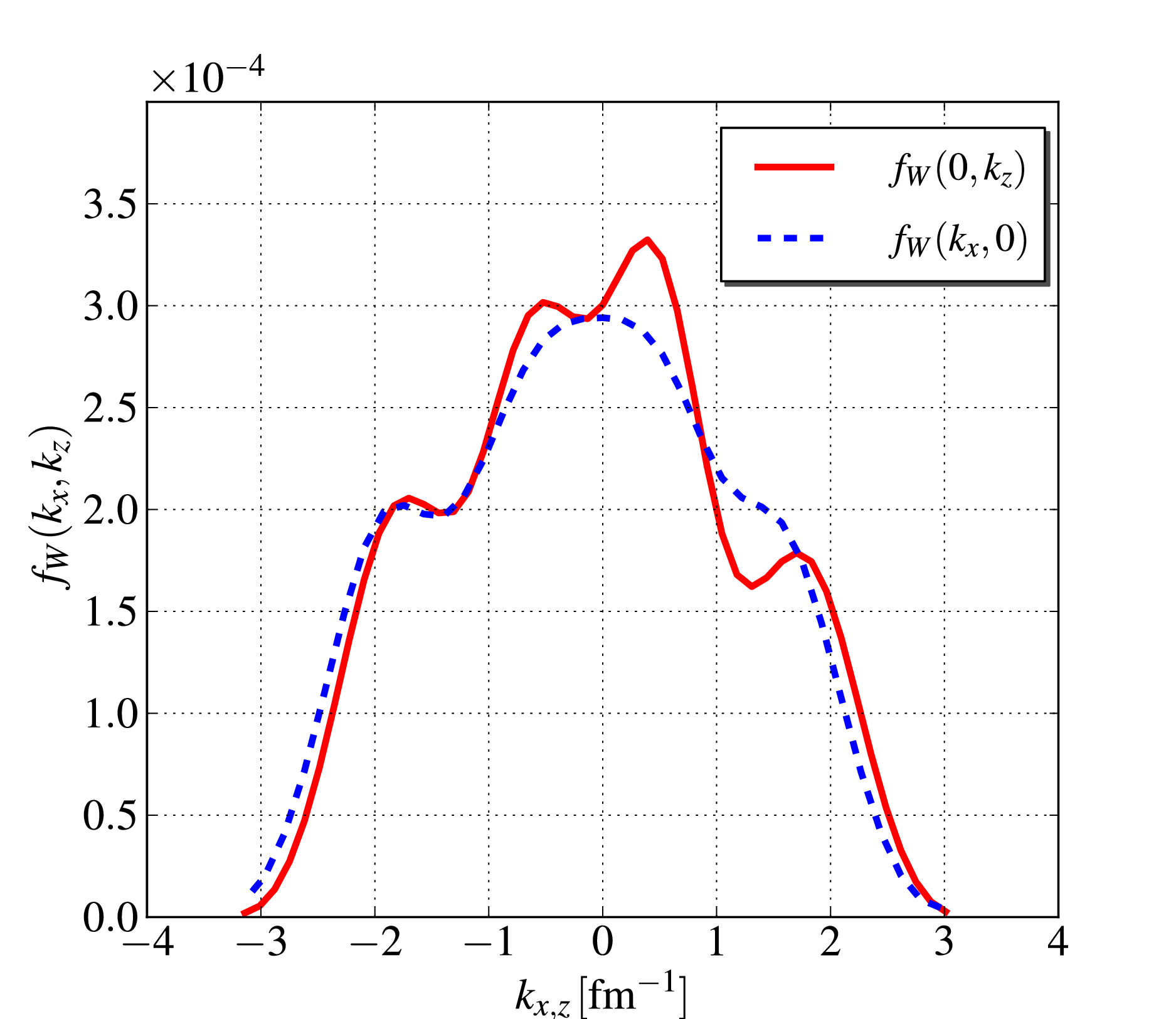}
\includegraphics*[width=8.6cm]{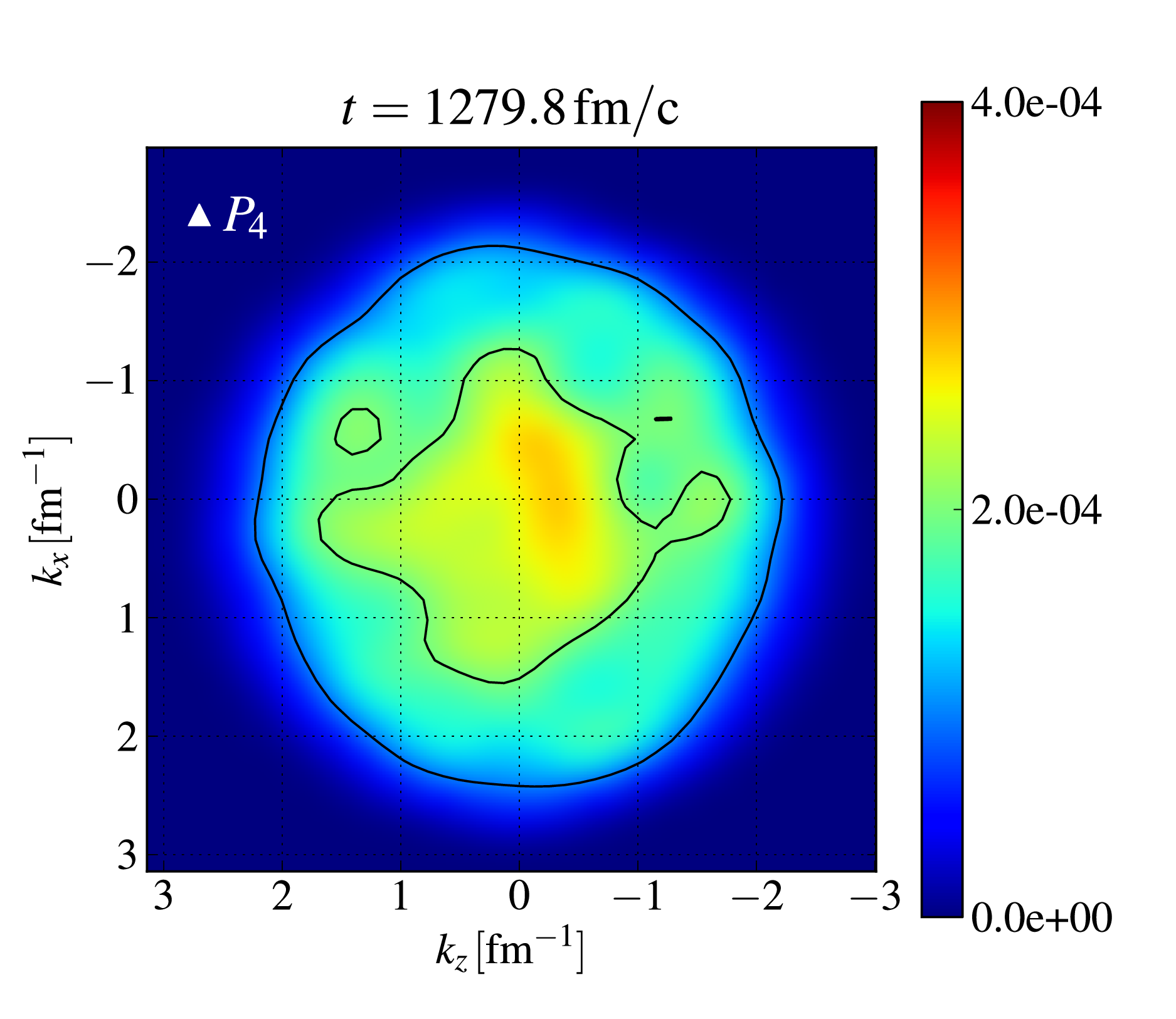}
\includegraphics*[width=8.6cm]{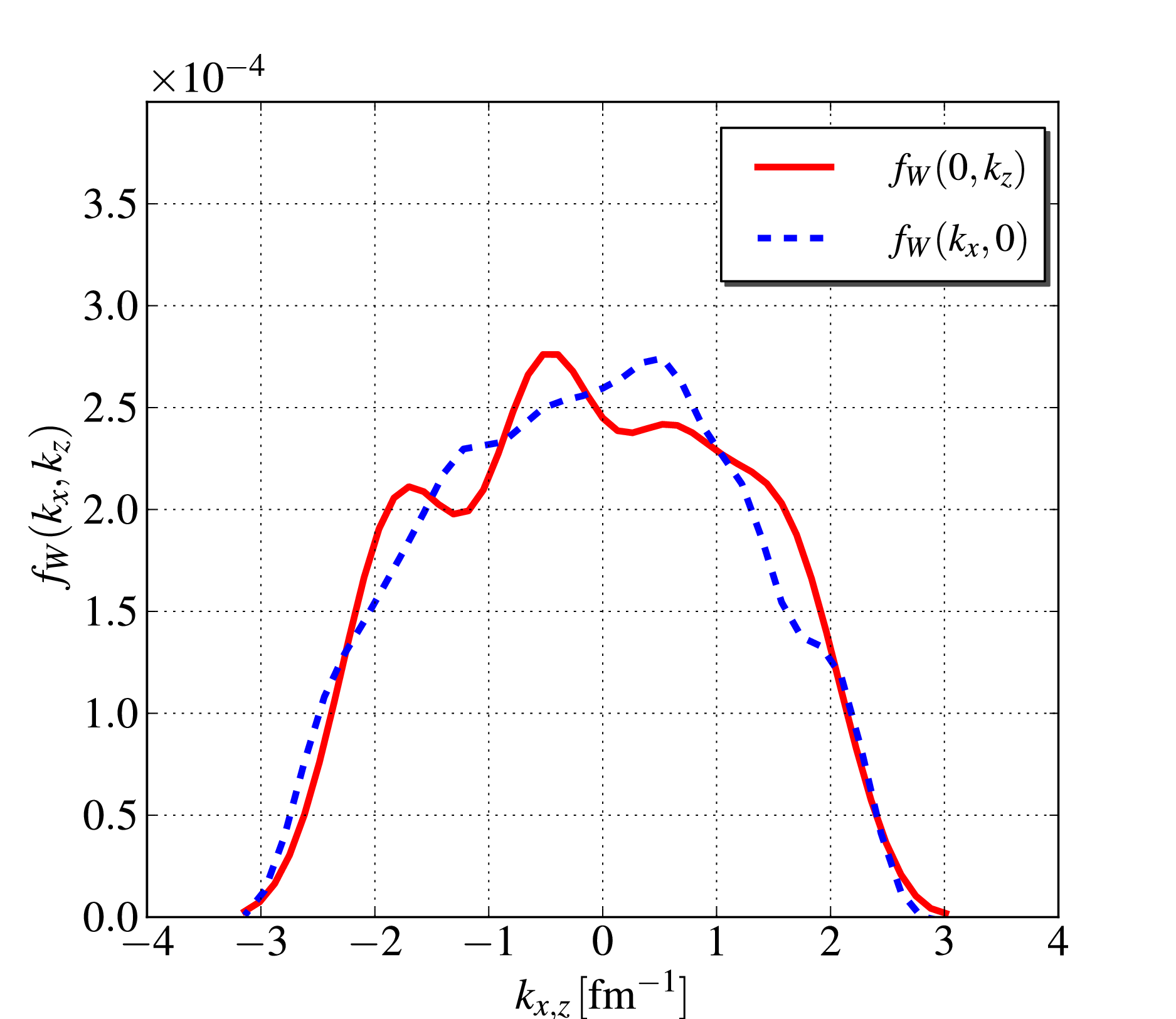}
\caption{\label{fig:ZrSnMomentB} (color online)
Same as Figure \ref{fig:ZrSnMomentA} but for the selected
(inner) points in Figure \ref{fig:ratio2d}.}
\end{figure*}

\begin{figure*}[!htbp]
\includegraphics*[width=8.6cm]{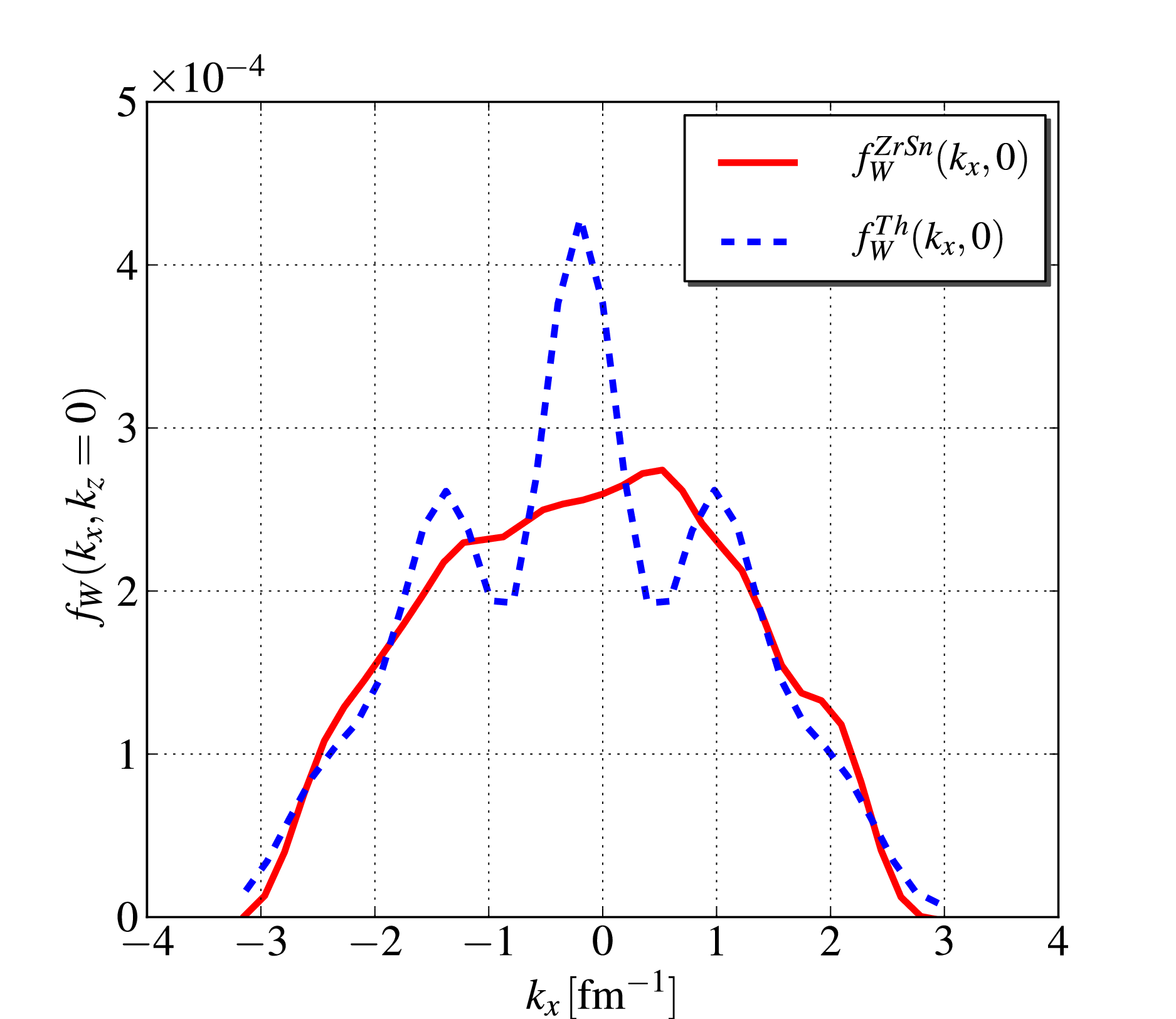}
\includegraphics*[width=8.6cm]{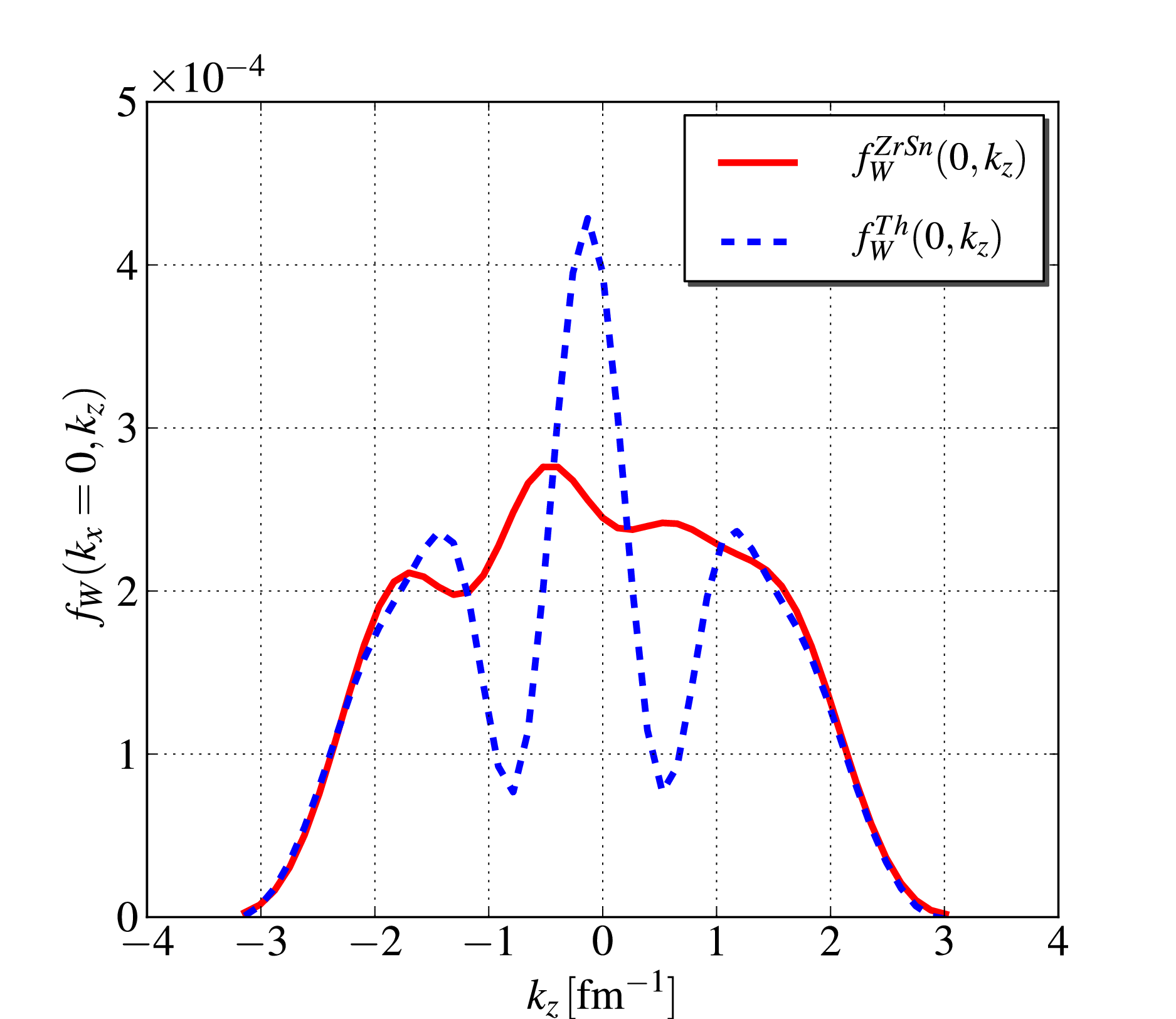}
\caption{\label{fig:Th230ZrSn} (color online)
Slices along $k_x$ (left) and $k_z$ (right) through
$f^{(2)}_\mathrm{W}(k_x,k_z)$ are shown for the merged $^{96}Zr+$$^{132}Sn$
system (red) at $t=1279.8$\:fm/c in the point $P_4$ from Figure
\ref{fig:ratio2d} in comparison with the ground state of $^{230}Th$
(blue).}
\end{figure*}

Figure \ref{fig:fusion} shows the global eccentricity $\varepsilon(t)$, the
internal kinetic energy $E_{int}(t)$, and the fragment distance $d(t)$ for both
reactions.  The distance oscillates after the minimum reached on first impact.  This
indicates that both cases describe a fused compound state.  The eccentricity
$\varepsilon(t)$ follows the oscillations of the distance, reflecting
a continuing vivid interaction with the spatial deformation. This
indicates that we are far from equilibration. The intrinsic kinetic energy
grows initially and soon levels off, leaving small residual oscillations about
a constant mean value. This mean intrinsic energy is, of course, larger for
the higher-energy collision ($E_{c.m.}=50$\:MeV). The example demonstrates nicely
that one needs a couple of observables to conclude on equilibration.  One may
be tempted to take the constant $E_{int}(t)$ as indicator of a thermalized
state. The still large values of eccentricity and the oscillations thereof
prove clearly that we are rather in a situation of substantial coherent
oscillations of the compound system.  In this context, it is to be remembered
that the energy stored in the collective motion of the compound system is
subtracted in the evaluation of $E_{int}(t)$.

The demonstrated behavior of the observables was checked up to $t=4000$\:fm/c,
twice the time span shown in Figure \ref{fig:fusion}. The pattern carried on
unchanged also for these longer times.

\subsection{$^{96}$Zr+$^{132}$Sn}
\label{sec:ZrSn}

As an example for a much heavier nuclear system we present fusion of
$^{96}$Zr+$^{132}$Sn achieved with a center-of-mass energy of
$E_{c.m.}=250$\:MeV and impact parameter $b=2$\:fm. The reaction between the
neutron-rich $^{132}$Sn nucleus and $^{96}$Zr was already studied in TDHF with a
focus on barrier heights and widths of the heavy-ion potential as well as capture
cross sections \cite{Oberacker}.

It was not possible in the present analysis to calculate the distance $d(t)$
between the fragments as it was done in the $^{16}$O+$^{16}$O fusion
scenario. The numerical algorithm selecting the spatial expectation values for
a two-body system was not able to detect two distinct objects during the whole
calculation and in this asymmetric system a simple symmetric division of the
grid was not possible. In Figure \ref{fig:ZrSn} we therefore use the
expectation value $Q_{20}\equiv\langle \hat{Q}_{20}\rangle$ of the quadrupole
operator $\hat{Q}_{20}$ to visualize the global geometry of the reaction.
Again, large values indicate separated fragments and low values a compound
stage.  It is obvious from the figure that the reaction ends in a compound
nucleus.  The overall trends of intrinsic kinetic energy and eccentricity are to
some extent similar to the results in Figure \ref{fig:fusion}. However, the
final eccentricity is much smaller, still maintaining some small
oscillations. This indicates a better thermalization than seen for
$^{16}$O+$^{16}$O, which is no surprise because the single-particle phase space
is much larger for the heavier system. The trend of the intrinsic energy does
also differ in detail. There seem to be two stages of growth, a fast
initial rise on the way to the compound stage and a slower, but steady,
growth up to 1000 fm/c. This indicates that some thermalization processes
and energy transport from deformation energy to kinetic energy is still
going on. After 1000 fm/c we again see a rather constant $E_\mathrm{int}$
as seems to be typical for energetic compound nuclei.

Figure \ref{fig:ratio2d} shows the local ratio $R^{(2)}(\mathbf{r},t)$ as
defined in Eq. (\ref{eq:ratio}) in the reaction plane at
$t=1279.8$\:fm/c.  The surface region is distinguished by large values coming
close to the Maxwellian reference values while much smaller ratios are seen
inside.  In order to illuminate these results, we have a closer look at the more
detailed  momentum distributions at four selected points 
indicated in Figure \ref{fig:ratio2d}. Figure \ref{fig:ZrSnMomentA} shows
results for two the outer points at the surface. The first point $(P_1)$ is
taken at approximately half the maximum value of $f_\mathrm{W}$. The
distribution is very similar to a Gaussian overlayed by a slight
asymmetry. The next point $(P_2)$ shows a more pronounced asymmetric
shape. Moving further to the inner points $(P_3,P_4)$ reviewed in Figure
\ref{fig:ZrSnMomentB} the momentum distributions differ substantially from
Gaussians and come closer to the idea of a Fermi distribution, although
heavily overlayed by quantum shell oscillations.

An similar analysis of the momentum distribution at other points
near the nuclear center yields similar results. The inner region of
the merged systems seems to stay rather ``cold`` during the reaction.

The strong quantum mechanical shell oscillations hinder a fit of the
distribution functions shown in Figure \ref{fig:ZrSnMomentA} and
\ref{fig:ZrSnMomentB} to a Fermi function from which one eventually could read
off an estimate for the system's temperature distribution. Therefore we compare
the $^{96}$Zr+$^{132}$Sn system to be assumed "hot`` with the "cold" analogue
of this system. Figure \ref{fig:Th230ZrSn} shows the momentum
distribution of the $^{96}$Zr+$^{132}$Sn compound system at the point $P_4$
indicated in Figure \ref{fig:ZrSnMomentB}. This is compared with the result from
the prolate ground state of $^{230}$Th. The $^{96}$Zr+$^{132}$Sn system
consists of $p=90$ protons and $n=138$ neutrons. $^{230}$Th shares the same
proton number with two additional neutrons. The ground state nucleus shows
huge, fully developed shell oscillations. Compared to these, the remaining
quantum oscillations in the "hot" compound state become rather small. The
disappearance of quantum shell effects is a major thermalization effect
\cite{Bra97aB,Bra93}. The occupation of high momentum components, however,
which would also be expected for hot systems remains insignificant.  This is due
to the fact that the nucleus is an open system from which high energy particles
escape, constantly depleting the high-momentum parts of the distribution. 
This explains why a fit
to Fermi distributions failed. A measure of temperature may be deduced from the
suppression of the shell oscillations, but this analysis is blurred by the
large thermal fluctuations in the momentum-space density. For the time being,
the eccentricity remains the cleanest indicator of equilibration.

\section{Summary}

In this work we have analyzed from different perspectives the dynamics
of TDHF during various reactions including the nuclei $^{16}O$,
$^{96}Zr$, $^{132}Sn$, and $^{230}Th$ with various center-of-mass
energies and impact parameters. The key quantity of the analysis is the
Wigner distribution function which provides a detailed phase-space
picture of the quantum state.  As complementing quantities, we also
considered three more compact observables in terms of local
distributions: the ratio $R(\mathbf{r},t)$ of the weighted moments
(weight four and two) of the local momentum distribution described by
the Wigner function (i.e. integrating the Wigner function over
momentum space for fixed local position), the  eccentricity 
$\epsilon(\mathbf{r},t)$ of the local momentum distribution, and the
intrinsic excitation energy $E_\mathrm{intr}(\mathbf{r},t)$ as deduced
from the kinetic energy density.

General properties of the Wigner distribution were discussed first for
stationary states. It shows oscillations which stem from the quantum
shell oscillations of the underlying single-particle states.  We also
looked at the Husimi function derived from the Wigner function by
some phase-space smoothing. The latter indeed provides a cleaner and
more intuitive picture. We find, however, that the shell oscillations
are much reduced in the dynamical scenarios of heavy-ion collisions,
allowing us to continue the dynamical studies with the
Wigner function alone.

We have visualized the collision process through snapshots of a 2D cut
of the 6D Wigner function. This shows that the two initially separated
phase space blobs never fully merge, even at the compound stage.
The distributions of the emerging fragments acquire a strong asymmetry
in momentum space and nicely show  the phase space rotations
associated with the remaining octupole oscillations of the final
fragments. 

The moment ratio $R$ was intended as a means to
compare the shape of the TDHF-Wigner distribution with a Maxwellian
distribution corresponding to thermal equilibrium. We find for all reaction
parameters that the moment ratio remains below the Maxwellian
reference value, which means that thermalization could not be asserted
in this observable. The reason is that that the high-momentum tail of
the actual distribution is immediately depleted by particle emission.
This exemplifies the fact that a true equilibrium state is hard to
establish in an open system.

The eccentricity $\varepsilon$ turned out to be a more useful
indicator. It grows dramatically in the initial phase of the reaction
and relaxes to lower values quickly after the compound state and then
remains oscillating about some finite value. This means that the final
relaxation to a thermal state is probably underestimated in mere TDHF.
Similar patterns are shown by 
the 
intrinsic kinetic energy $E_{int}$.
The resulting ``asymptotic'' value of  $E_{int}$ depends
strongly on the initial conditions, e.g., growing with the initial
collision energy.

We conclude that although TDHF includes dissipation owing to
single-particle viscosity, which acts strongly in the initial phase of
reactions, there is no evidence for complete equilibration.

\section*{Acknowledgment}
This work was supported by the Frankfurt Center for Scientific
Computing and by the BMBF under Contracts No. 06FY9086 and 06ER9063. We gratefully acknowledge
support by the Frankfurt Center for Scientific Computing.

\end{document}